\documentclass[12pt]{iopart}
\usepackage{etoolbox}
\usepackage{slashed}
\usepackage{iopams}

\expandafter\let\csname equation*\endcsname\relax

\expandafter\let\csname endequation*\endcsname\relax

\usepackage{amsmath,bm}
\usepackage{amsthm}
\usepackage{amssymb}
\usepackage{graphicx}
\usepackage[bb=boondox]{mathalfa}
\usepackage{hhline}
\usepackage{mathrsfs}
\usepackage{braket}
\usepackage{caption}
\usepackage[colorlinks,citecolor=blue,urlcolor=blue,linkcolor=blue]{hyperref}
\usepackage{xcolor}
\usepackage{soul}
\usepackage{caption}
\usepackage{subcaption}

\makeatletter
\newcommand{\mainmatter}{%
  \setcounter{footnote}{0}%
  \patchcmd{\@makefntext}{\fnsymbol}{\arabic}{}{}%
  \patchcmd{\@thefnmark}{\fnsymbol}{\arabic}{}{}%
  \def\@makefnmark{\textsuperscript{\arabic{footnote}}}%
}
\makeatother
\begin{document}

\title{Smearing out contact terms in ghost-free infinite derivative quantum gravity}

\author{Ulrich K. Beckering Vinckers$^{1,2,*}$, \'{A}lvaro de la Cruz-Dombriz$^{1,3}$ and Anupam Mazumdar$^2$}
\address{$^1$Cosmology and Gravity Group, Department of Mathematics and Applied Mathematics,
University of Cape Town, Rondebosch 7701, Cape Town, South Africa}
\address{$^2$Van Swinderen Institute, University of Groningen, 9747 AG Groningen, The Netherlands}
\address{$^3$Departamento de F\'{i}sica Fundamental, Universidad de Salamanca, 37008 Salamanca, Spain}
    
\eads{$^*$ bckulr002@myuct.ac.za}

\begin{abstract}
In the context of ghost-free infinite derivative gravity, we consider the single graviton exchange between two spinless particles, a spinless particle and a photon, or between a spinless particle and a spin-1/2 particle. To this end, we compute the gravitational potential for the three aforementioned cases and derive the $\mathcal{O}(G\hbar^2)$ correction that arises at the linearised level. In the local theory, it is well-known that such a correction appears in the form of a Dirac delta function. Here, we show that this correction is smeared out for the nonlocal theory and, in contrast to the local theory, takes on non-zero values for a non-zero separation between the two particles. In the case of the single graviton exchange between a spinless particle and a spin-1/2 particle, we also compute the $\mathcal{O}(G\hbar)$ correction that arises in the non-static case within the non-relativistic approximation and show that it is finite in the nonlocal theory.
\end{abstract}

\maketitle
\mainmatter

\section{Introduction}
An extension of General Relativity (GR) allows us to construct infinite derivative theories of gravity (IDG)~\cite{Krasnikov:1987yj,Tomboulis:1997gg, Siegel:2003vt,Biswas:2005qr,Biswas:2011ar,Modesto:2011kw}. This is because a long-range massless graviton mediates the gravitational interaction. Hence, infinitely many derivatives of curvature invariants can be constructed with the help of covariant form factors, e.g., ${\cal F}(\Box)$ where $\Box=g_{\mu\nu}\nabla^{\mu}\nabla^{\nu}$ and $\mu,\nu\in\{0,1,2,3\}$. We also note that a $(-,+,+,+)$ signature is used throughout this work. IDG theories introduce a nonlocal gravitational interaction with a nonlocal scale, which has been constrained to be $0.004$~eV from the current bounds on the Newtonian potential ~\cite{Edholm:2016hbt}. A nonlocal gravitational action can be recast using quadratic curvature terms, including the ${\mathcal F}(\Box)$ form factors, such that it modifies the ultraviolet (UV) behavior of gravity
while recovering the infrared (IR) behavior at low energies\footnote{Nonlocal IR modifications of gravity can also be constructed in terms of the form factor ${\mathcal F}(\Box^{-1})$, see \cite{Deser:2007jk,Woodard:2014iga,Conroy:2014eja}. Here, we will focus on the UV modification of the propagator.}. The IDG gravitational action contains infinitely many derivatives, and to avoid introducing ghost degrees of freedom, one takes specific analytic functions which recover the local limit of GR smoothly~\cite{Edholm:2016hbt,Biswas:2011ar}. The inclusion of torsion has been implemented in Refs.~\cite{delaCruz-Dombriz:2018aal,delaCruz-Dombriz:2019tge}, although here we will not consider torsion.

It was shown that a very particular form of nonlocality arises in string field theory, see \cite{Siegel:2003vt,deLacroix:2017lif} and, in particular, p-adic strings \cite{Freund:2017aqf,FREUND1987191}. In \cite{Abel:2019zou}, it was suggested that nonlocal field theories might even arise naturally by discarding the higher modes of a first-quantised string theory within a particle approximation, which makes the nonlocal theory a natural consequence for fields in the UV~\cite{Buoninfante:2018mre}.
The ghost-free IDG can resolve the cosmological singularity \cite{Biswas:2005qr,Biswas:2011ar,Biswas:2012bp,Biswas:2010zk,
Kumar:2020xsl,Kolar:2021qox,Koshelev:2018rau} and the anisotropic Kasner singularity \cite{Kumar:2021mgc}. Also, the ghost-free IDG has been shown to resolve a variety of point-like singularities \cite{Biswas:2005qr,Biswas:2011ar,Frolov:2015bia,Frolov:2015bta,Frolov:2015usa,Frolov:2016xhq,Buoninfante:2018xiw,Buoninfante:2018rlq,Buoninfante:2018stt,Buoninfante:2018xif,Boos:2018bxf,Vinckers:2022the,Burzilla:2020utr} and there is motivation~\cite{Koshelev:2017bxd,Buoninfante:2019swn,Maggio:2020jml} to suggest that astrophysical black hole singularities could be resolved.
 Exact solutions for the ghost-free IDG have been constructed \cite{Biswas:2005qr,Kolar:2021rfl,Kolar:2020ezu,Kolar:2020max,Kolar:2023gqi,Kilicarslan:2019njc,Dengiz:2020xbu,Kolar:2021uiu} and there is a good hint that nonlocal field theories may ameliorate renormalisable properties of gravity \cite{Tomboulis:1997gg, Modesto:2011kw,Talaganis:2014ida, Modesto:2017hzl,Tomboulis:2015esa,Boos:2021jih,Boos:2021lsj}. Furthermore, there exists extensions of ghost-free IDG in $1+2$ dimensions~\cite{Mazumdar:2018xjz}, in de Sitter~\cite{Biswas:2016etb} and maximally symmetric spacetimes~\cite{SravanKumar:2019eqt}. 

The present work highlights the importance of the $\mathcal{O}(G\hbar^2)$ correction that arises when considering the single graviton exchange between two particles in IDG, see~\cite{Barker:1966zz,michael1979advanced} for the case of GR. This issue has been neglected so far in the nonlocal gravity literature. In the context of GR, $\mathcal{O}(G\hbar^2)$ corrections in the form of Dirac delta functions occur and are dubbed as \textit{contact terms} \cite{Barker:1966zz}. They are usually repulsive and arise as the result of analytic terms that appear in the scattering amplitude. Even the Coulomb potential in quantum electrodynamics has a similar behavior; see, for example~\cite {michael1979advanced}. However, it is essential to emphasise that for a non-zero separation between the two particles, these contact terms give a zero contribution to the potential energy. It is also worth pointing out that, for the static case of two spinless particles in the context of GR, the contact term that appears is independent of the particles' masses~\cite{Barker:1966zz,michael1979advanced}.

Here, we wish to examine the effect of nonlocality on the contact terms that arise in the single graviton exchange either between two massive spinless particles, between a massive spinless particle and a photon\footnote{At higher-order in GR~\cite{Bjerrum-Bohr:2014zsa} and in the case of sixth-order gravity~\cite{Accioly:2016qeb,Accioly:2016lzp} the light bending scattering potential has been calculated at one-loop and tree-level, respectively. We also note that the light bending angle in IDG was obtained in~\cite{Feng:2017vqd} while~\cite{Draper:2020knh} discusses the scattering of spinless particles in the context of IDG.}, or between a massive spinless particle and a massive spin-1/2 particle.  We will show that such terms become smeared out in these three cases. In addition, these terms can be repulsive for a non-zero separation between the two particles, providing a deeper understanding of how the nonlocal gravitational interaction naturally ameliorates the UV aspects of spacetime. Although our analysis is perturbative, it is nevertheless helpful to understand how, in future, we may be able to probe such effects by resorting to entanglement tests of quantum gravity, see~\cite{Bose:2017nin, Marshman:2019sne,Bose:2022uxe,Vinckers:2023grv}, provided the scale of nonlocality is close to the experimental parameters, i.e., within $\sim {\cal O}(10^{-4} -10^{-3})~{\rm m}$.

While we only consider the three scattering scenarios mentioned above we note that \cite{Barker:1966zz} considers, in addition to these three scenarios, the interaction between a photon and a spin-1/2 particle as well as the interaction between two spin-1/2 particles. Regarding higher-derivative gravity, in~\cite{deBrito:2020wmp} non-static interactions at tree-level between two spinless particles and also between two spin-1/2 particles are considered.
\section{Linearised IDG}
In what follows, we consider a Minkowski background and therefore write $\Box=\eta^{\mu\nu}\partial_\mu\partial_\nu$ where $\eta_{\mu\nu}$ is the Minkowski metric in Cartesian coordinates. Here, the perturbed metric is related to the full metric through $h_{\mu\nu}=(g_{\mu\nu}-\eta_{\mu\nu})/\kappa$ where $\kappa^2=16\pi G$ in natural units. The quadratic part of the Einstein-Hilbert (EH) action contains $h^{\mu\nu}\Box h_{\mu\nu}$, $h\partial_\mu\partial_\nu h^{\mu\nu}$, $h^{\mu\nu}\partial_\mu\partial_\alpha h_\nu\phantom{}^\alpha$ and $h\Box h$ terms in its Lagrangian, and a nonlocal modification of said action can be obtained by inserting infinite derivative operators into each of these four terms\footnote{We note that this is a specific case of the action given in \cite{Biswas:2011ar} which allows for a fifth term in the gravitational Lagrangian.}. The requirement that the conservation equation be satisfied leads to only one of these operators being independent, resulting in the following quadratic infinite derivative action~\cite{Biswas:2011ar,Biswas:2013cha}
\begin{align}\label{eq:quadratic_IDG_action}
&S_{\text{G}}=\frac14\int\text{d}^4x\Big[h_{\mu\nu}\mathcal{F}\left(\Box\right)\Box h^{\mu\nu}+2h\mathcal{F}\left(\Box\right)\partial_\mu\partial_\nu h^{\mu\nu}-2h^{\mu\nu}\mathcal{F}\left(\Box\right)\partial_\mu\partial_\alpha h_\nu\phantom{}^\alpha-h\mathcal{F}\left(\Box\right)\Box h\Big]\,.
\end{align}
The propagator for the above action, i.e., the free nonlocal theory, has been derived before~\cite{Biswas:2011ar,Biswas:2013kla} and may be obtained by finding the Green's function for the equations of motion. In the de Donder gauge, it is given by
\begin{align}\label{eq:Green's_function_linearised_IDG}
G_{\mu\nu}\phantom{}^{\alpha\beta}(x,x')&=-\left(\delta_\mu^\alpha\delta_\nu^\beta+\delta_\nu^\alpha\delta_\mu^\beta-\eta_{\mu\nu}\eta^{\alpha\beta}\right)\frac1{2\left(2\pi\right)^4}\int\text{d}^4k\frac{{\text e}^{ik\cdot\left(x-x'\right)}}{\mathcal{F}\left(-k^2\right)k^2}\,.
\end{align}
For derivations of the Green's function~\eqref{eq:Green's_function_linearised_IDG} and the quadratic IDG action \eqref{eq:quadratic_IDG_action} the reader is directed to \ref{sec:derive_greens}. From \eqref{eq:Green's_function_linearised_IDG} it is clear that the quadratic IDG action \eqref{eq:quadratic_IDG_action} does not admit any additional degrees of freedom compared to GR, and is therefore ghost-free\footnote{It is worth pointing out that, while the theory is ghost-free at tree-level, it has been shown in~\cite{Shapiro:2015uxa} that unphysical states may arise at one-loop. Nevertheless, this is beyond the scope of the present work since we only consider tree-level computations}, provided that ${\mathcal F}(\Box)$ is analytic with no zeros\footnote{Although not written explicitly, when the form factor is analytic with no zeros, we make use of the following prescription
\begin{align}
G_{\mu\nu}\phantom{}^{\alpha\beta}(x,x')=-\left(\delta_\mu^\alpha\delta_\nu^\beta+\delta_\nu^\alpha\delta_\mu^\beta-\eta_{\mu\nu}\eta^{\alpha\beta}\right)\frac1{2\left(2\pi\right)^4}\lim_{\epsilon\rightarrow0^+}\int\text{d}^4k\frac{{\text e}^{ik\cdot(x-x')}}{\mathcal{F}\left(-k^2\right)\left(k^2-i\epsilon\right)}\,,
\end{align}
when evaluating the integral in Eq.~\eqref{eq:Green's_function_linearised_IDG}.
}. In other words, all of the poles in the integral \eqref{eq:Green's_function_linearised_IDG} are contained in the factor of $1/k^2$. An example of $\mathcal{F}(\Box)$ satisfying this requirement is the exponential of an entire function. The most straightforward choice is 
\begin{align}\label{eq:form_factor_choice}
{\mathcal F}(\Box)= {\text e}^{-\ell^2\Box}\,,
\end{align}
where $\ell>0$ is the so-called \textit{length scale of nonlocality}. The choice \eqref{eq:form_factor_choice} makes the UV properties of the theory well-behaved and recovers GR in the IR~\cite{Siegel:2003vt}, i.e., in the limit $\ell \rightarrow 0$. Such a form factor is also motivated by the low-energy limit of string theory~\cite{Abel:2019zou}. For the coupling between matter and gravity, we use the standard prescription which gives us a matter action of the form
\begin{align}\label{eq:matter_action}
S_{\text m}=\frac\kappa2\int\text{d}^4xh^{\mu\nu}T_{\mu\nu}\,,
\end{align}
where $T_{\mu\nu}$ denotes the energy-momentum tensor expanded up to zeroth-order in $\kappa$. We also make use of the gauge-fixing action \cite{Vinckers:2023grv}
\begin{align}\label{eq_gauge_fixing_action}
S_{\text{GF}}=-\frac12\int\text{d}^4x&\left(\partial_\mu h^\mu\phantom{}_\nu-\frac12\partial_\nu h\right)\mathcal{F}(\Box)\left(\partial_\alpha h^{\alpha\nu}-\frac12\partial^\nu h\right)\,.
\end{align}
The total action under consideration is now
\begin{align}\label{eq:gauged_fixed_action}
S_{\text{total}}=S_{\text G}+S_{\text m}+S_{\text{GF}}\,.
\end{align}

In the present work, we consider three matter fields: massive spinless fields, Maxwell fields and massive spin-1/2 fields. In the following, we shall discuss the energy-momentum tensors on a momentum basis for each case.

\begin{itemize}

\item \textit{Massive spinless field}. By expanding the spinless field in terms of creation and annihilation operators, then writing the energy-momentum tensor in terms of these operators and finally taking excited states to be relativistically normalised, one can obtain the following
\begin{align}\label{eq:T_mu_nu_massive_kets}
\bra{p'}T_{\mu\nu}(x)&\ket{p}={\text e}^{ix\cdot (p-p')}\left[p_\mu p'_\nu+p_\nu p'_\mu-\eta_{\mu\nu}(p\cdot p'+m^2)\right]\,.
\end{align}
The $p$ and $p'$ are the incoming and outgoing momenta, respectively, and $m$ is the mass. We also note that the on-shell conditions $p^2=p'^2=-m^2$ are taken to be satisfied.

\item \textit{Maxwell field}. In this case, by expanding first the electromagnetic four-potential in terms of creation and annihilation operators, then writing the corresponding energy-momentum tensor in terms of these operators, and finally taking excited states to be relativistically normalised, one obtains the following
\begin{align}\label{eq:maxwell_field_energy_momentum_tensor_final}
&\bra{\tau',p'}T_{\mu\nu}(x)\ket{p,\tau}={\text e}^{ix\cdot(p-p')}\epsilon^\alpha\phantom{}_\tau(\bm p)\epsilon^\beta\phantom{}_{\tau'}(\bm p')\big[\eta_{\alpha\beta}(p_\mu p'_\nu+p_\nu p'_\mu)-\eta_{\beta\nu}p_\mu p'_\alpha-\eta_{\alpha\nu}p_\beta p'_\mu\nonumber\\
&-\eta_{\beta\mu}p_\nu p'_\alpha-\eta_{\alpha\mu}p_\beta p'_\nu+p\cdot p'\left(\eta_{\alpha\mu}\eta_{\beta\nu}+\eta_{\alpha\nu}\eta_{\beta\mu}-\eta_{\mu\nu}\eta_{\alpha\beta}\right)+\eta_{\mu\nu}p_\beta p'_\alpha\big]\,.
\end{align}
We note that the on-shell conditions $p^2=p'^2=0$ for the photon are satisfied. In addition, we have used $\epsilon^{\alpha\tau}$ to denote the four polarisation vectors of the photon~\cite{tong_lecture} and the indices $\tau$ and $\tau'$ to denote the polarisation of the initial and final states, respectively. In addition, these polarisation vectors satisfy the condition $\epsilon^{\mu\alpha}\epsilon_\mu\phantom{}^\beta=\eta^{\alpha\beta}$. Furthermore, we take $\epsilon^{\mu1}$ and $\epsilon^{\mu2}$ to be transverse to the momentum of the photon while taking $\epsilon^{\mu3}$ to be longitudinal.
\item\textit{Massive spin-1/2 field}. By expanding the Dirac fields in terms of creation and annihilation operators and then writing the energy-momentum tensor in terms of said operators, one can obtain \cite{michael1979advanced}
\begin{align}\label{eq:spinor_field_energy_momentum_tensor_final}
\bra{s',p'}&T_{\mu\nu}(x)\ket{p,s}=\frac14{\text e}^{ix\cdot(p-p')}\bar u_{s'}(p')\left[\gamma_\mu(p_\nu+p'_\nu)+(p_\mu+p'_\mu)\gamma_\nu\right]u_s(p)\,,
\end{align}
where $s,s'\in\{1,2\}$ denote the spin state and
\begin{align}\label{eq:u_s_def}
u_s(p)=\begin{pmatrix}\sqrt{p^0-\bm\sigma\cdot\bm p}\xi_s\\ \sqrt{p^0+\bm\sigma\cdot\bm p}\xi_s\end{pmatrix}\,,
\end{align}
where the $\xi_s$ are two-component numerical spinors and satisfy the normalisation condition $\xi^\dagger_{s'}\xi_s=\delta_{s's}$. It is also noted that $\bar u_s(p)=u^\dagger_s(p)\gamma^0$ and we make use of the Weyl representation for the $\gamma^\mu$ matrices following~\cite{Peskin:1995ev} (see Eq.~\eqref{eq:gamma_matrices}).
\end{itemize}

In \ref{sec:quantum_energy_momentum_tensors_derivations}, we review the derivations for the quantum energy-momentum tensor expressions given in Eqs.~\eqref{eq:T_mu_nu_massive_kets}, \eqref{eq:maxwell_field_energy_momentum_tensor_final} and \eqref{eq:spinor_field_energy_momentum_tensor_final}.
\section{Scattering via graviton exchange}
For the single graviton exchange between two particles, $A$ and $B$, we denote the initial and final states for the interaction as $\ket i=\ket{i_A,i_B}$ and $\ket f=\ket{f_A,f_B}$,
respectively. The $\ket{i_{A,B}}$ and $\ket{f_{A,B}}$ are used to denote, respectively, the initial and final states of the individual particles, with each consisting of a single particle excitation. The $S$-matrix element for the single graviton exchange between the two particles is given by~\footnote{Since we are considering distinguishable particle scattering, the creation and annihilation operators contained in $T_A^{\mu\nu}$ do not act on the $\ket{i_B}$ and $\ket{f_B}$ states, and analogously for $T_B^{\mu\nu}$ and the $\ket{i_A}$ and $\ket{f_A}$ states.}:
\begin{align}\label{eq:S_matrix_element_3}
\bra fS-1&\ket i=-\frac{i\kappa^2}{2}\int\text{d}^4x\text{d}^4x'\bra{f_A}T_A^{\mu\nu}(x)\ket{i_A}G_{\mu\nu\alpha\beta}\left(x,x'\right)\bra{f_B}T_B^{\alpha\beta}\left(x'\right)\ket{i_B}\,.
\end{align}
A derivation of Eq.~\eqref{eq:S_matrix_element_3} can be found in \ref{sec:s_matrix_derivation}. The scattering amplitude $\mathcal{A}$ is extracted from the $S$-matrix element via 
\begin{align}\label{eq:extract_scattering_amplitude}
\bra fS-1\ket i=i\mathcal{A}(2\pi)^4\delta^{(4)}(P_{fi})\,,
\end{align}
where $P_{fi}$ denotes the difference between the system's total initial and final momenta, see~\cite{michael1979advanced,tong_lecture}. In the following, we will use $p$ and $p'$ to denote, respectively, the incoming and outgoing momenta of the $A$ particle while using $q$ and $q'$ for that of the $B$ particle. For derivations of the scattering amplitudes discussed in the following sections, the reader is directed to \ref{sec:scattering_amplitude_derivations}.
\subsection{Two spinless particles}
To evaluate the $S$-matrix element~\eqref{eq:S_matrix_element_3}, we use the following initial and final states: $\ket{i_A}=\ket p$, $\ket{f_A}=\ket{p'}$, $\ket{i_B}=\ket q$ and $\ket{f_B}=\ket{q'}$. In addition, we denote the mass of the $A$ and $B$ particles by $m_A$ and $m_B$, respectively. The inner product $\bra{f_A}T_A^{\mu\nu}(x)\ket{i_A}$ is now given by the right-hand side of~\eqref{eq:T_mu_nu_massive_kets} with $m$ replaced by $m_A$ while $\bra{f_B}T_B^{\mu\nu}(x)\ket{i_B}$ is also given by the right-hand side of~\eqref{eq:T_mu_nu_massive_kets} but with $p$, $p'$ and $m$ replaced with $q$, $q'$ and $m_B$, respectively. By evaluating the $S$-matrix element~\eqref{eq:S_matrix_element_3}, we can  extract the following scattering amplitude
\begin{align}\label{eq:amplitude_spinless}
\mathcal{A}=\frac{\kappa^2}{\mathcal{F}(-t)t}\left[2\left(p\cdot q\right)^2-p^2q^2+tp\cdot q\right]\,,
\end{align}
where  $t=(p'-p)^2$ is one of the three so-called \textit{Mandelstam invariants} \cite{Mandelstam:1958xc,michael1979advanced} with the other two being $(p-q')^2$ and $(p+q)^2$. In what follows, we also use the \textit{graviton transfer momentum} given by $k=p'-p=q-q'$.

At this point, we use the notation $E_p=p^0$ and $E_q=q^0$ and consider the centre-of-mass (CM) frame, i.e., $\bm q=-\bm p$ and $\bm q'=-\bm p'$. As the result of energy conservation, we have $p'^0=p^0$ and $q'^0=q^0$ and thus $t=\bm k^2$. With this in mind, we divide the last expression~\eqref{eq:amplitude_spinless} by minus four times the product of the two particles' energies, i.e., $-4E_pE_q$, since we have considered relativistically normalised states, and take its inverse Fourier transform. Doing this yields the following gravitational potential
\begin{align}\label{eq:spinless_potential_general}
V(\bm r)&=-\frac{GE_pE_q}{c^4}\left[1+\left(4+\frac{E_q}{E_p}+\frac{E_p}{E_q}\right)\frac{\bm p^2c^2}{E_pE_q}+\frac{\bm p^4c^4}{E_p^2E_q^2}\right]f_1(\bm r)+\frac{G\hbar^2}{c^2}\left(1+\frac{\bm p^2c^2}{E_pE_q}\right)f_0(\bm r)\,,
\end{align}
where we define
\begin{align}\label{eq:f_n_definition}
f_n(\bm r)=4\pi\int\frac{\text{d}^3k}{(2\pi)^3}\frac{{\text e}^{i\bm k\cdot\bm r}}{\mathcal{F}(-\bm k^2)\bm k^{2n}}\,.
\end{align}
We note that we have used $\kappa^2=16\pi G$, which is true in natural units, and then introduced $\hbar$ and $c$ through dimensional analysis. For the local case, i.e., $\mathcal{F}(\Box)=1$, we have $f_1(\bm r)=1/r$ whereas $f_0(\bm r)=4\pi\delta^{(3)}(\bm r)$ and thus Eq.~\eqref{eq:spinless_potential_general} reduces to the well-known quantum GR result which was obtained in \cite{Barker:1966zz} (see Eq.~(14) therein).

We now wish to examine the potential~\eqref{eq:spinless_potential_general} for the case where the form factor is given by Eq.~\eqref{eq:form_factor_choice}. For such a case, the gravitational potential is
\begin{align}\label{eq:spinless_potential}
V(\bm r)=-\frac{GE_pE_q}{c^4r}&\left[1+\left(4+\frac{E_q}{E_p}+\frac{E_p}{E_q}\right)\frac{\bm p^2c^2}{E_pE_q}+\frac{\bm p^4c^4}{E_p^2E_q^2}\right]\text{erf}\left(\frac{r}{2\ell}\right)\nonumber\\
&+\frac{G\hbar^2}{2\sqrt\pi c^2\ell^3}\left(1+\frac{\bm p^2c^2}{E_pE_q}\right){\text e}^{-\frac{r^2}{4\ell^2}}\,,
\end{align}
where $\text{erf}(x)$ denotes the so-called \textit{error function} \cite{andrews1998special}. Eq. \eqref{eq:spinless_potential} gives us the linearised gravitational potential for the IDG theory and its $\mathcal{O}(G\hbar^2)$ correction when the nonlocal form factor is given by Eq.~\eqref{eq:form_factor_choice}. On the one hand, for the specific case where the momenta of the two particles are along the same axis, the $\mathcal{O}(G\hbar^0)$ term has been computed before in~\cite{Vinckers:2023grv} for testing the entanglement between the two masses~\cite{Bose:2017nin}. Such a result is the nonstatic generalisation of the potential found in \cite{Biswas:2011ar,Marshman:2019sne} and smears out the singular nature contained in the factor of $1/r$ through the error function. On the other hand, the $\mathcal{O}(G\hbar^2)$ term is computed for the first time here and we now wish to compare it to its GR counterpart. The GR result may be obtained by taking the local limit $\ell\rightarrow0$ in~\eqref{eq:spinless_potential}. In such a limit, the error function tends to unity while the exponential behaves as ${\text e}^{-(x^2+y^2+z^2)/(4\ell^2)}/\ell^3\rightarrow(4\pi)^{3/2}\delta^{(3)}(\bm r)$. Thus, the local limit of Eq.~\eqref{eq:spinless_potential} gives us the quantum GR result. The $\mathcal{O}(G\hbar^2)$ correction for the quantum GR case is a Dirac delta function, i.e., a contact term, and is, therefore, exactly zero for a non-zero separation. Remarkably, the $\mathcal{O}(G\hbar^2)$ correction in the nonlocal potential \eqref{eq:spinless_potential} is smeared out and is non-zero for a non-zero separation.

As a special case of \eqref{eq:spinless_potential}, let us consider the static case by setting $\bm p=0$. For such a case, we have $E_p=m_Ac^2$ and $E_q=m_Bc^2$ and Eq.~\eqref{eq:spinless_potential} reduces to
\begin{align}\label{eq:spinless_potential_static}
&V_{\text{static}}(\bm r)=-\frac{Gm_Am_B}{r}\text{erf}\left(\frac{r}{2\ell}\right)+\frac{G\hbar^2{\text e}^{-\frac{r^2}{4\ell^2}}}{2\sqrt\pi c^2\ell^3}\,.
\end{align}
As is also true for the static contact term of GR, see, for example, \cite{michael1979advanced}, the $\mathcal{O}(G\hbar^2)$ correction given in \eqref{eq:spinless_potential_static} is independent of the particle masses and is repulsive. Finally, it is worth considering the limit as $r\rightarrow 0$ in Eq.~\eqref{eq:spinless_potential_static}. For such a consideration, the error function behaves as $r/(\ell\sqrt\pi)$ which gives $V_{\text{static}}\rightarrow -Gm_Am_B/(\sqrt\pi\ell) +G\hbar^2/(2\sqrt{\pi} c^2\ell^3)$. Therefore, when
\begin{align}\label{eq:spin_0_ell_r_to_0}
\ell \sim \frac{\hbar}{c \sqrt{2m_Am_B}}\,,
\end{align}
the static gravitational potential vanishes in the limit $r\rightarrow 0$. 

We emphasise that we have only considered the linearised regime and there are higher-order corrections in $G$ that are not considered here and would contribute to the potential. Nevertheless, it is worth noting that the condition in Eq.~\eqref{eq:spin_0_ell_r_to_0} results in the first-order in $G$ contributions to the static gravitational potential canceling in the limit as $r\rightarrow0$.
\subsection{Photon and spinless particle} For our initial and final states of the photon ($A$), we use $\ket{i_A}=\ket{p,s}$ and $\ket{f_A}=\ket{p',s'}$.  Here, $s,s'\in\{1,2\}$ correspond to the transverse polarisations of the incoming and outgoing photon states. For the initial and final spinless particle ($B$) states, we use the same initial and final states as in the previous section, i.e., $\ket{i_B}=\ket q$, and $\ket{f_B}=\ket{q'}$. The inner product $\bra{f_A}T_A^{\mu\nu}(x)\ket{i_A}$ is now given by the right-hand side of Eq.~\eqref{eq:maxwell_field_energy_momentum_tensor_final} while $\bra{f_B}T_B^{\mu\nu}(x)\ket{i_B}$ is given by the right-hand side of Eq.~\eqref{eq:T_mu_nu_massive_kets} but instead written in terms of $q$, $q'$ and $m_B$. We also note that in the following we shall use the indices $i,j\in\{1,2,3\}$ to denote the spatial components of tensors. By considering the CM frame, we arrive at the following $S$-matrix element using Eq.~\eqref{eq:S_matrix_element_3}
\begin{align}\label{eq:scattering_matrix_photon}
\bra fS-1\ket i=i\kappa^2&(2\pi)^4\delta^{(4)}(p-p'+q-q')\epsilon^i\phantom{}_s(\bm p)\epsilon^j\phantom{}_{s'}(\bm p')\nonumber\\
&\times\bigg\{\delta_{ij}\left[2(p\cdot q)^2+tp\cdot q+\frac{tq^2}{2}\right]-k_ik_j\left(2p\cdot q+q^2\right)\bigg\}\frac{1}{\mathcal{F}(-t)t}\,,
\end{align}
where $k$ is the transfer momentum as before. It is worth noting that to obtain Eq.~\eqref{eq:scattering_matrix_photon} we have used the fact that both $\epsilon^i\phantom{}_s(\bm p)p_i$ and $\epsilon^i\phantom{}_{s'}(\bm p')p'_i$ are zero since these polarisation vectors are the transverse ones. In addition, it follows that $\epsilon^i\phantom{}_s(\bm p)q_i$ and $\epsilon^i\phantom{}_{s'}(\bm p')q'_i$ are also zero since we are in the CM frame. Using Eq.~\eqref{eq:extract_scattering_amplitude} we can extract the scattering amplitude, which we denote as $\mathcal{A}_{ss'}$ for this case, from the last expression~\eqref{eq:scattering_matrix_photon}. In order to obtain the gravitational potential, we first average the scattering amplitude $\mathcal{A}_{ss'}$ over its transverse polarisations. We therefore require the summations of $\epsilon^i\phantom{}_s(\bm p)\epsilon_{is'}(\bm p')$ and $\epsilon^i\phantom{}_s(\bm p)\epsilon^j\phantom{}_{s'}(\bm p')k_ik_j$ over $s$ and $s'$. For our choice of polarisation vectors, we follow \cite{Barker:1966zz} and take both\footnote{Here, we use $\bm\epsilon_s(\bm p)$ and $\bm\epsilon_{s'}(\bm p')$ to denote the transverse polarisation vectors whose components are $\epsilon^i\phantom{}_s(\bm p)$, and $\epsilon^j\phantom{}_{s'}(\bm p')$, respectively.} $\bm\epsilon_1(\bm p)$ and $\bm\epsilon_1(\bm p')$ to be the unit vector that is orthogonal to both the incoming and outgoing three-momenta of the photon, i.e., $\bm\epsilon_1(\bm p)=\bm\epsilon_1(\bm p')=\left(\bm p\times\bm p'\right)/|\bm p\times\bm p'|$. On the other hand, we take the second of the transverse polarisation vectors of the incoming and outgoing photon momenta to be
\begin{align}
\bm\epsilon_2(\bm p)=\frac{\bm p\times\bm\epsilon_1(\bm p)}{|\bm p|}\,,\ \ \ \ \ \bm\epsilon_2(\bm p')=\frac{\bm p'\times\bm\epsilon_1(\bm p')}{|\bm p'|}\,,
\end{align}
respectively. Using the fact that the angle $\theta$ between the incoming and outgoing photon momenta is related to the transfer momentum through $\bm k^2=2\bm p^2(1-\cos\theta)$ in the CM frame and that the transfer momentum is orthogonal to $\bm\epsilon_1(\bm p)$ and $\bm\epsilon_1(\bm p')$, one finds the following expressions
\begin{align}
\frac12\sum_{s,s'}\epsilon^i\phantom{}_s(\bm p)\epsilon_{is'}(\bm p')&=1-\frac{\bm k^2}{4\bm p^2}\,,\label{eq:spin_averaged_sum_1}\\
\frac12\sum_{s,s'}\epsilon^i\phantom{}_s(\bm p)\epsilon^j\phantom{}_{s'}(\bm p')k_ik_j&=\frac{\bm k^2}2\left(1-\frac{\bm k^2}{4\bm p^2}\right)\label{eq:spin_averaged_sum_2}\,.
\end{align}
Therefore, by making use of Eqs.~\eqref{eq:spin_averaged_sum_1} and~\eqref{eq:spin_averaged_sum_2}, one can obtain the following expression for the scattering amplitude averaged over the transverse polarisations
\begin{align}\label{eq:scattering_amplitude_photon}
\frac12\sum_{s,s'}\mathcal A_{ss'}&=\frac{\kappa^2}{\mathcal{F}(-\bm k^2)\bm k^2}\left(2-\frac{\bm k^2}{2\bm p^2}\right)\left(p\cdot q\right)^2\,.
\end{align}

As before, we denote $E_p=p^0$ and $E_q=q^0$ and note that $E_p=|\bm p|$ in natural units. By dividing Eq.~\eqref{eq:scattering_amplitude_photon} by minus four times the product of the two particles' energies, i.e., $-4E_pE_q$ and taking the inverse Fourier transform, one obtains the gravitational potential for the interaction between a photon and a massive spinless particle in the context of the quantum IDG theory at the linearised level
\begin{align}\label{eq:potential_photon_general}
V(\bm r)&=-\frac{2GE_pE_q}{c^4}\left(1+\frac{2E_p}{E_q}+\frac{E_p^2}{E_q^2}\right)f_1(\bm r)+\frac{G\hbar^2}{c^2}\left(1+\frac{E_p}{2E_q}+\frac{E_q}{2E_p}\right)f_0(\bm r)\,.
\end{align}
We also note that we have once again introduced $\hbar$ and $c$ through dimensional analysis and that the local case, i.e., $\mathcal{F}(\Box)=1$, reduces Eq.~\eqref{eq:potential_photon_general} to the quantum GR result that was obtained in \cite{Barker:1966zz} (see Eq.~(42) therein). When the form factor is given by Eq.~\eqref{eq:form_factor_choice}, we obtain
\begin{align}\label{eq:potential_photon_1}
V(\bm r)&=-\frac{2GE_pE_q}{c^4r}\left(1+\frac{2E_p}{E_q}+\frac{E_p^2}{E_q^2}\right)\text{erf}\left(\frac{r}{2\ell}\right)+\frac{G\hbar^2{\text e}^{-\frac{r^2}{4\ell^2}}}{2\sqrt\pi c^2\ell^3}\left(1+\frac{E_p}{2E_q}+\frac{E_q}{2E_p}\right)\,.
\end{align}
Again, we note that the $\mathcal{O}(G\hbar^2)$ correction in the ghost-free IDG theory is finite and smeared out with a width proportional to the length scale of nonlocality, as opposed to the Dirac delta function appearing in the local case.
As was done in the previous section for the case of two massive spinless particles, we examine the potential energy~\eqref{eq:potential_photon_1} in the limit as $r\rightarrow0$. In such a limit, we find that for a nonlocal length scale of
\begin{align}\label{eq:ell_photon}
\ell \sim \frac{c}{\omega\sqrt8}\,,    
\end{align}
where $\omega=E_p/\hbar$ is the angular frequency of the photon, the potential associated with the single graviton exchange between a photon and a massive spinless particle vanishes in the context of the quantum IDG theory. It is worth noting that \eqref{eq:ell_photon} is independent of the mass of the spinless particle.\\

\subsection{Spin-1/2 particle and a spinless particle}
The procedure followed here to obtain the $S$-matrix element for this interaction is similar to what was done in the previous section, except that the $A$ particle is now a massive spin-1/2 particle instead of a photon. More specifically we use $\ket{i_A}=\ket{p,s}$ and $\ket{f_A}=\ket{p',s'}$, respectively, for the initial and final states of the spin-1/2 particle where $s$ and $s'$ denote its spin states. The inner product $\bra{f_A}T_A^{\mu\nu}(x)\ket{i_A}$ appearing in the expression~\eqref{eq:S_matrix_element_3} is now given by the right-hand side of Eq.~\eqref{eq:spinor_field_energy_momentum_tensor_final}. Hence, the expression for the $S$-matrix element is now found to be
\begin{align}\label{eq:scattering_matrix_spin_half}
\bra fS-1\ket i=\frac{i\kappa^2}{4}(2\pi)^4&\delta^{(4)}(p-p'+q-q')\bar u_{s'}(p')\nonumber\\
&\times\left[4\slashed qp\cdot q-2\slashed pq^2+t\slashed q-\slashed k\left(2p\cdot q+q^2+\frac{t}{2}\right)\right]\frac{u_s(p)}{\mathcal{F}(-t)t}\,,
\end{align}
where $\slashed p=\gamma^\mu p_\mu$ and so on. In what follows, we consider the CM frame and the non-relativistic approximation. Thus, we only consider the scattering amplitude up to second-order in three-momentum terms. In addition, and as was similarly done for the case where a photon and a spinless particle exchange a graviton, we extract the scattering amplitude $\mathcal{A}_{ss'}$ and perform a spin-averaged sum. Keeping in mind the non-relativistic approximation, we first note that
\begin{align}
\bar u_{s'}(p')\gamma^0u_s(p)&\approx\xi^\dagger_{s'}\left[2p^0+\frac{(\bm\sigma\cdot\bm k)(\bm\sigma\cdot\bm p)}{2p^0}\right]\xi_s\,,\label{eq:u_0_u}\\
\bar u_{s'}(p')\gamma^iu_s(p)&\approx\xi^\dagger_{s'}\left(2p^i+\bm\sigma\cdot\bm k\sigma^i\right)\xi_s\,.\label{eq:u_i_u}
\end{align}
We also note that $\left(\bm\sigma\cdot\bm k\right)\left(\bm\sigma\cdot\bm p\right)=i\bm\sigma\cdot(\bm k\times\bm p)-\bm k^2/2$ and now find
\begin{align}\label{eq:sum_scattering_amplitude_spin}
\frac12\sum_{s,s'}\mathcal{A}_{ss'}\approx\frac{\kappa^2\psi^\dagger}{\mathcal{F}(-\bm k^2)\bm k^2}\bigg[2\left(p\cdot q\right)^2&-p^2q^2+i\bm\sigma\cdot\left(\bm k\times\bm p\right)\left(p^0q^0+\frac{3\left(q^0\right)^2}{4}\right)\nonumber\\
&-\bm k^2\left(p^0q^0+\frac{3\left(q^0\right)^2}{8}\right)\bigg]\psi\,,
\end{align}
where we define the normalised two-component spinor $\psi=2^{-1/2}(\xi_1+\xi_2)$. The gravitational potential, which we treat as acting on the two-component spinor $\psi$, is now found to be
\begin{align}\label{eq:spin_1/2_potential}
V(\bm r)&=-Gm_Am_B\left[1+\frac{\bm p^2}{m_Am_Bc^2}\left(4+\frac{3m_A}{2m_B}+\frac{3m_B}{2m_A}\right)\right]f_1(\bm r)\nonumber\\
&-\frac{G\hbar}{c^2}\left(1+\frac{3m_B}{4m_A}\right)\bm\sigma\cdot\left[\nabla f_1(\bm r)\times\bm p\right]+\frac{G\hbar^2}{c^2}\left(1+\frac{3m_B}{8m_A}\right)f_0(\bm r)\,.
\end{align}
In the local case, since once again we would have $f_1(\bm r)=1/r$, and thus $\nabla(1/r)=-\bm r/r^3$, as well as $f_0(\bm r)=4\pi\delta^{(3)}(\bm r)$, Eq.~\eqref{eq:spin_1/2_potential} would reduce to the quantum GR result (see Eq.~(23) in \cite{Barker:1966zz}).

As was done in the previous two cases, let us consider the case where the form factor is given by Eq.~\eqref{eq:form_factor_choice}. For such a case, Eq.~\eqref{eq:spin_1/2_potential} yields the following potential
\begin{align}\label{eq:spin_1/2_potential_erf}
&V(\bm r)=-\frac{Gm_Am_B}{r}\left[1+\frac{\bm p^2}{m_Am_Bc^2}\left(4+\frac{3m_A}{2m_B}+\frac{3m_B}{2m_A}\right)\right]\text{erf}\left(\frac{r}{2\ell}\right)\nonumber\\
&+\frac{G\hbar}{c^2}\left(1+\frac{3m_B}{4m_A}\right)\left[\frac{\text{erf}\left(\frac{r}{2\ell}\right)}{r}-\frac{\text{e}^{-\frac{r^2}{4\ell^2}}}{\sqrt\pi\ell}\right]\frac{\bm\sigma\cdot\bm L}{r^2}+\frac{G\hbar^2{\text e}^{-\frac{r^2}{4\ell^2}}}{2\sqrt\pi c^2\ell^3}\left(1+\frac{3m_B}{8m_A}\right)\,,
\end{align}
where $\bm L=\bm r\times\bm p$ is the angular momentum. It is straightforward to verify that the $\mathcal{O}(G\hbar)$ correction in Eq. \eqref{eq:spin_1/2_potential_erf} is finite in the limit as $r\rightarrow0$ whereas, in the case of GR, it diverges. By considering the static case, Eq.~\eqref{eq:spin_1/2_potential_erf} gives us
\begin{align}\label{eq:spin_1/2_potential_erf_static}
&V(\bm r)=-\frac{Gm_Am_B}{r}\text{erf}\left(\frac{r}{2\ell}\right)+\frac{G\hbar^2{\text e}^{-\frac{r^2}{4\ell^2}}}{2\sqrt\pi c^2\ell^3}\left(1+\frac{3m_B}{8m_A}\right)\,.
\end{align}
It is worth pointing out that the last expression~\eqref{eq:spin_1/2_potential_erf_static} coincides with Eq.~\eqref{eq:spinless_potential_static}, which is the static potential in the case of two spinless particles, provided that the mass of the spin-1/2 particle is much larger than that of the spinless particle, i.e., $m_A\gg m_B$. By taking the limit $r\rightarrow0$, we find that the potential~\eqref{eq:spin_1/2_potential_erf_static} vanishes for
\begin{align}\label{eq:spin_1/2_ell_r_to_0}
\ell\sim\frac{\hbar}{c\sqrt{2m_Am_B}}\sqrt{1+\frac{3m_B}{8m_A}}\,,
\end{align}
which agrees with Eq.~\eqref{eq:spin_0_ell_r_to_0} when $m_A\gg m_B$.
\section{Behavior for various form factors}
\begin{figure}
\includegraphics[width=0.7\columnwidth]{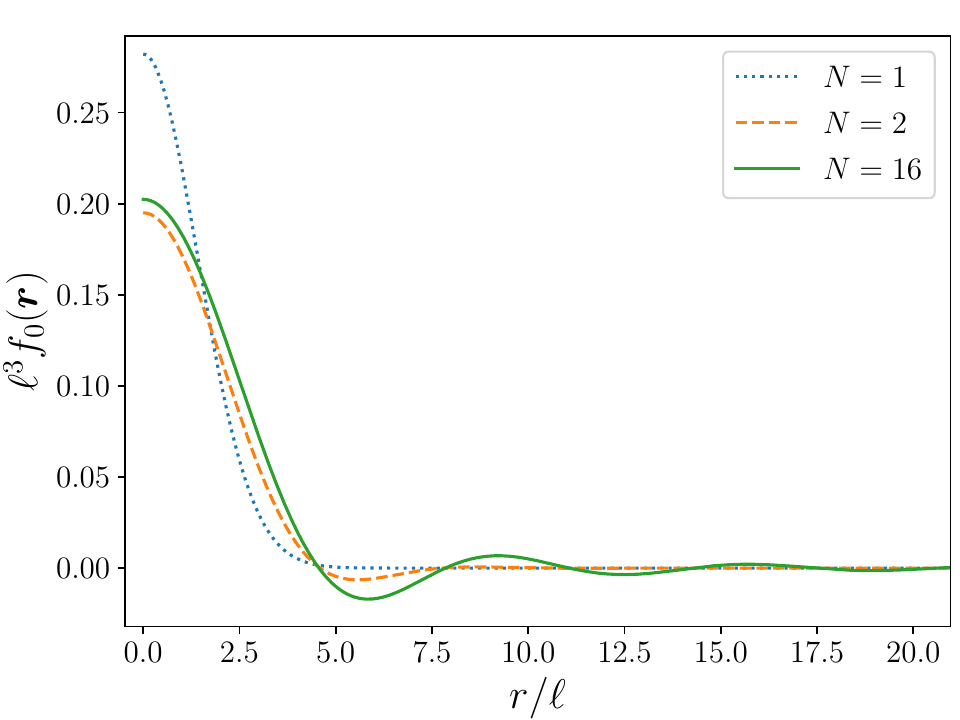}
\caption{Function $f_0(\bm r)$ describing the behaviour of the $\mathcal{O}(G\hbar^2)$ correction in the nonlocal gravitational potentials~\eqref{eq:spinless_potential_general},~\eqref{eq:potential_photon_general} and~\eqref{eq:spin_1/2_potential} when the form factor considered is of the form given in Eq.~\eqref{eq:choice_form_factor_N}. The dotted blue, dashed orange and solid green curves correspond to $N=1$, $N=2$ and $N=16$, respectively. Although the $\mathcal{O}(G\hbar^2)$ correction is finite for all three cases considered, the $N=2,\,16$ cases allow for the correction to be attractive for some values of $r/\ell$, while in the $N=1$ case such a correction is repulsive for all distances.}
\label{fig:f_0_plot}
\end{figure}

Herein we examine how the $\mathcal{O}(G\hbar^2)$ correction that appears in Eqs.~\eqref{eq:spinless_potential_general},~\eqref{eq:potential_photon_general} and~\eqref{eq:spin_1/2_potential} behaves for the following choice of a form factor:
\begin{align}\label{eq:choice_form_factor_N}
\mathcal{F}(\Box)={\text e}^{-\ell^{2N}\Box^N}\,,
\end{align}
where we take $N\in\mathbb{Z}^+$. Such a form factor has been considered in~\cite{Edholm:2016hbt,Frolov:2015usa} for classical IDG. In the present work, we have so far considered the $N=1$ case only and now wish to consider additional values of $N$. In Fig.~\ref{fig:f_0_plot} we plot $f_0(\bm r)$, which is defined in Eq.~\eqref{eq:f_n_definition},
when the nonlocal form factor is given by Eq.  \eqref{eq:choice_form_factor_N}
with $N=1$, $N=2$ or $N=16$. This figure shows that in the $N=1$ case, $f_0(\bm r)$ is positive. Such a function can take on negative values in the $N=2$ and $N=16$ cases. Therefore, we conclude that there exist ghost-free IDG form factors that allow for the $\mathcal{O}(G\hbar^2)$ correction to be attractive for certain values of $r/\ell$ while being repulsive for others.

\section{Conclusions}
In this manuscript, we have computed and examined the single graviton exchange between two massive spinless particles, between a massive spinless particle and a photon, and between a massive spinless particle and a massive spin-$1/2$ particle in the context of the quantum IDG theory. For the first time, we computed the nonlocal analogue of the contact term that arises in the local theory and showed that it becomes smeared out. In contrast to the contact term of the local theory, which is identically zero for a non-zero separation between the two particles, the corresponding $\mathcal{O}(G\hbar^2)$ correction within the nonlocal theory admits non-zero values. We also showed that, in the limit as the separation between the two particles tends to zero, there are values for the nonlocal length scale that allow for the potential to vanish. Lastly, we note that we also computed the $\mathcal{O}(G\hbar)$ correction that arises for the nonstatic interaction between a spinless particle and a spin-1/2 particle in the nonlocal theory and showed that it is indeed finite as opposed to divergent, which is what occurs in the local theory.
\section{Acknowledgments} UKBV acknowledges financial support from the National Research Foundation of South Africa, Grant number PMDS22063029733, from the University of Cape Town Postgraduate Funding Office and the University of Groningen. AdlCD acknowledges support from BG20/00236 action (MCINU, Spain), NRF Grant CSUR23042798041, CSIC Grant COOPB23096 and Grant PID2021-122938NB-I00 funded by MCIN/AEI/10.13039/501100011033 and by {\it ERDF A way of making Europe}.
\appendix
\section{Derivation of the IDG quadratic action and Green's function}\label{sec:derive_greens}
In this section we review the derivation of the IDG quadratic action given in Eq.~\eqref{eq:quadratic_IDG_action} as well as the associated Green's function in the de Donder gauge given in Eq.~\eqref{eq:Green's_function_linearised_IDG}. Such expressions have been derived before in Refs.~\cite{Biswas:2011ar,Biswas:2013cha}. We start with reviewing the derivation of Eq.~\eqref{eq:quadratic_IDG_action} in~\ref{sec:derive_action} before moving on to Eq.~\eqref{eq:Green's_function_linearised_IDG} in~\ref{sec:greens}.
\subsection{Derivation of Eq.~\eqref{eq:quadratic_IDG_action}}\label{sec:derive_action}
Let us start with the EH action which is give by~\cite{Hilbert1915}
\begin{align}
S_{\text{EH}}=\frac{1}{\kappa^2}\int\text{d}^4x\sqrt{-g}R\,,
\end{align}
where $R$ is the Ricci scalar. For perturbations around Minkowski spacetime, the EH action may be approximated as follows
\begin{align}\label{eq:approx_EH}
S_{\text{EH}}\approx S_{\text{EH}}\big|_{g=\eta}-&\kappa\int\text{d}^4x\frac{\delta S_{\text{EH}}}{\delta g^{\mu\nu}}\bigg|_{g=\eta}\left(h^{\mu\nu}-\kappa h^\mu\phantom{}_\alpha h^{\alpha\nu}\right)\nonumber\\
&+\frac{\kappa^2}2\int\text{d}^4x\text{d}^4x'\frac{\delta^2S_{\text{EH}}}{\delta g^{\alpha\beta}(x')\delta g^{\mu\nu}(x)}\bigg|_{g=\eta}h^{\alpha\beta}(x')h^{\mu\nu}(x)\,,
\end{align}
where $h_{\mu\nu}=\left(g_{\mu\nu}-\eta_{\mu\nu}\right)/\kappa$ as before. We also note that the $S_{\text{EH}}$ and its functional derivatives appearing on the right-hand side of Eq.~\eqref{eq:approx_EH} are evaluated using the Minkowski metric. Furthermore, no boundary terms are considered. At this point, we note that both $S_{\text{EH}}$ and $\delta S_{\text{EH}}/\delta g^{\mu\nu}=\sqrt{-g}\left(R_{\mu\nu}-R g_{\mu\nu}/2\right)/\kappa^2$ vanish when evaluated using the Minkowski metric since the Ricci tensor is zero in such a case. Therefore, the first two terms on the right-hand side of \eqref{eq:approx_EH} vanish and we only need to consider the third. To this end, one can compute
\begin{align}\label{eq:delta_S_squared}
\kappa^2 h^{\alpha\beta}(x')h^{\mu\nu}(x)\frac{\delta^2S_{\text{EH}}}{\delta g^{\alpha\beta}(x')\delta g^{\mu\nu}(x)}\bigg|_{g=\eta}=\frac{1}{2}&h^{\alpha\beta}(x')h^{\mu\nu}(x)\big[\left(\eta_{\mu\alpha}\eta_{\nu\beta}-\eta_{\mu\nu}\eta_{\alpha\beta}\right)\Box-2\eta_{\mu\beta}\partial_{\nu}\partial_\alpha\nonumber\\
&+\eta_{\alpha\beta}\partial_\mu\partial_\nu+\eta_{\mu\nu}\partial_\alpha\partial_\beta\big]\delta^{(4)}(x-x')\,.
\end{align}
Substituting Eq.~\eqref{eq:delta_S_squared} into Eq.~\eqref{eq:approx_EH} gives
\begin{align}\label{eq:quadratic_EH}
S_{\text{EH}}\approx\frac14\int\text{d}^4x\left(h^{\mu\nu}\Box h_{\mu\nu}-h\Box h+2h\partial_\mu\partial_\nu h^{\mu\nu}-2h^\mu\phantom{}_\beta\partial_\mu\partial_\nu h^{\beta\nu}\right)\,.
\end{align}
An infinite derivative modification of Eq.~\eqref{eq:quadratic_EH} can be constructed as follows
\begin{align}\label{eq:S_G_no_constraint}
S_{\text{G}}=\frac14\int\text{d}^4x\left[h^{\mu\nu}\mathcal{F}_1(\Box)\Box h_{\mu\nu}-h\mathcal{F}_2(\Box)\Box h+2h\mathcal{F}_3(\Box)\partial_\mu\partial_\nu h^{\mu\nu}-2h^\mu\phantom{}_\beta\mathcal{F}_4(\Box)\partial_\mu\partial_\nu h^{\beta\nu}\right]\,,
\end{align}
where the $\mathcal{F}_i(\Box)$ may contain infinitely many derivatives. Let us now consider the conservation equations $\partial^\nu\left(\delta S_{\text G}/\delta h^{\mu\nu}\right)=0$ and examine how the nonlocal operators are constrained. Taking the functional derivative of Eq.~\eqref{eq:S_G_no_constraint} with respect to $h^{\mu\nu}$ gives
\begin{align}\label{eq:eom_h}
\frac{\delta S_{\text G}}{\delta h^{\mu\nu}}=\frac12\left[\mathcal{F}_1(\Box)\Box h_{\mu\nu}-\eta_{\mu\nu}\mathcal{F}_2(\Box)\Box h+\eta_{\mu\nu}\mathcal{F}_3(\Box)\partial_\alpha\partial_\beta h^{\alpha\beta}+\mathcal{F}_3(\Box)\partial_\mu\partial_\nu h-2\mathcal{F}_4(\Box)\partial_\alpha\partial_{(\mu}h^\alpha\phantom{}_{\nu)}\right]\,.
\end{align}
Taking the partial derivative of Eq.~\eqref{eq:eom_h} leads to
\begin{align}\label{eq:conservation_eq}
2\partial^\nu\left(\frac{\delta S_{\text G}}{\delta h^{\mu\nu}}\right)=\left[\mathcal{F}_1(\Box)-\mathcal{F}_4(\Box)\right]\Box\partial_\nu h^\nu\phantom{}_\mu-\left[\mathcal{F}_2(\Box)-\mathcal{F}_3(\Box)\right]\Box \partial_\mu h+\left[\mathcal{F}_3(\Box)-\mathcal{F}_4(\Box)\right]\partial_\mu\partial_\alpha\partial_\beta h^{\alpha\beta}\,.
\end{align}
In order for Eq.~\eqref{eq:conservation_eq} to vanish, we require the three terms above to independently vanish. This leads to
\begin{align}
\mathcal{F}_1(\Box)=\mathcal{F}_2(\Box)=\mathcal{F}_3(\Box)=\mathcal{F}_4(\Box)\,,
\end{align}
and we are thus left with only one independent form factor which we denote as $\mathcal{F}(\Box)$. Substituting such a condition into Eq.~\eqref{eq:S_G_no_constraint} yields Eq.~\eqref{eq:quadratic_IDG_action}.

For the sake of completeness, let us also note an action at the non-linear level that leads to the quadratic action Eq.~\eqref{eq:quadratic_IDG_action}. To this end, let us consider the action
\begin{align}
S=\frac1{\kappa^2}\int\text{d}^4x\sqrt{-g}\left[R+R\mathcal{G}_1(\Box)R+R_{\mu\nu}\mathcal{G}_2(\Box)R^{\mu\nu}\right]\,,
\end{align}
where the $\mathcal{G}_1(\Box)$ and $\mathcal{G}_2(\Box)$ may contain infinitely many derivatives. By making use of the linearised Ricci tensor:
\begin{align}
R_{\mu\nu}\approx\frac{\kappa}2\left(\partial_\mu\partial_\alpha h^\alpha\phantom{}_\nu+\partial_\nu\partial_\alpha h^\alpha\phantom{}_\mu-\Box h_{\mu\nu}-\partial_\mu\partial_\nu h\right)\,,
\end{align}
one can write the following integrals up to second-order in $\kappa$
\begin{align}
&\int\text{d}^4x\sqrt{-g}R\mathcal{G}_1(\Box)R\approx\kappa^2\int\text{d}^4x\left[h^{\mu\nu}\mathcal{G}_1(\Box)\partial_\mu\partial_\nu\partial_\alpha\partial_\beta h^{\alpha\beta}+h\mathcal{G}_1(\Box)\Box^2h-2h^{\mu\nu}\mathcal{G}_1(\Box)\Box\partial_\mu\partial_\nu h\right]\,,\label{eq:R_squared_action}\\
&\int\text{d}^4x\sqrt{-g}R_{\mu\nu}\mathcal{G}_2(\Box)R^{\mu\nu}\approx\frac{\kappa^2}4\int\text{d}^4x\big[h^{\mu\nu}\mathcal{G}_2(\Box)\Box^2h_{\mu\nu}+2h^{\mu\nu}\mathcal{G}_2(\Box)\partial_\mu\partial_\nu\partial_\alpha\partial_\beta h^{\alpha\beta}\nonumber\\
&\hspace{3cm}+h\mathcal{G}_2(\Box)\Box^2h-2h^{\mu\nu}\mathcal{G}_2(\Box)\Box\partial_\mu\partial_\nu h-2h^\mu\phantom{}_\alpha\mathcal{G}_2(\Box)\Box\partial_\mu\partial_\nu h^{\alpha\nu}\big]\,,\label{eq:R_mu_nu_squared_action}
\end{align}
where we have ignored any boundary terms. By making use of Eqs.~\eqref{eq:R_squared_action} and~\eqref{eq:R_mu_nu_squared_action} together with Eq.~\eqref{eq:quadratic_EH} one finds
\begin{align}\label{eq:grouped_quadratic}
S\approx\frac14\int\text{d}^4x\big\{h_{\mu\nu}\left[1+\mathcal{G}_2(\Box)\Box\right]\Box h^{\mu\nu}-h\left[1-4\mathcal{G}_1(\Box)\Box-\mathcal{G}_2(\Box)\Box\right]\Box h+2h[1-\mathcal{G}_2(\Box)\Box\nonumber\\
-4\mathcal{G}_1(\Box)\Box]\partial_\mu\partial_\nu h^{\mu\nu}-2h^\mu\phantom{}_\beta\left[1+\mathcal{G}_2(\Box)\Box\right]\partial_\mu\partial_\nu h^{\beta\nu}+2h^{\mu\nu}\left[2\mathcal{G}_1(\Box)+\mathcal{G}_2(\Box)\right]\partial_\mu\partial_\nu\partial_\alpha\partial_\beta h^{\alpha\beta}\big\}\,.
\end{align}
Equating Eq.~\eqref{eq:grouped_quadratic} and Eq.~\eqref{eq:quadratic_IDG_action} yields~\cite{Biswas:2013cha}
\begin{align}\label{eq:covariant_form_factor_constraint}
-2\mathcal{G}_1(\Box)=\mathcal{G}_2(\Box)=\frac{\mathcal{F}(\Box)-1}\Box\,.
\end{align}
It follows then that the action\footnote{We note that, while Eq.~\eqref{eq:covariant_form_factor_constraint} in the present work coincides with Eq.~(87) of~\cite{Biswas:2013cha}, we were unable to derive the action given in Eq.~(88) of~\cite{Biswas:2013cha} and instead find Eq.~\eqref{eq:non_linear_action} of the present work. For a comparison of notation, note that $\mathcal{F}(\Box)$, $\mathcal{G}_1(\Box)$ and $\mathcal{G}_2(\Box)$ in the present work correspond to $a(\Box)$, $2\mathcal{F}_1(\Box)$ and $2\mathcal{F}_2(\Box)$ in~\cite{Biswas:2013cha}. In addition, the factor of $1/\kappa^2$ used here is instead a factor of 1/2 in~\cite{Biswas:2013cha}.}
\begin{align}\label{eq:non_linear_action}
S=\frac{1}{\kappa^2}\int\text{d}^4x\sqrt{-g}\left[R-R\left(\frac{\mathcal{F}(\Box)-1}{2\Box}\right)R+R_{\mu\nu}\left(\frac{\mathcal{F}(\Box)-1}{\Box}\right)R^{\mu\nu}\right]\,,
\end{align}
leads to Eq.~\eqref{eq:quadratic_IDG_action} in the linearised regime.
\subsection{Derivation of Eq.~\eqref{eq:Green's_function_linearised_IDG}}\label{sec:greens}
Let us first consider the matter action $S_{\text m}$ whose corresponding energy-momentum tensor can be obtained in the usual way:
\begin{align}\label{eq:energy_momentum_def}
T_{\mu\nu}=-\frac2{\sqrt{-g}}\frac{\delta S_{\text m}}{\delta g^{\mu\nu}}\,.
\end{align}
The matter action may be approximated as
\begin{align}\label{eq:approx_matter_action}
S_{\text m}\approx-\kappa\int\text{d}^4x\frac{\delta S_{\text m}}{\delta g^{\mu\nu}}\bigg|_{g=\eta}h^{\mu\nu}\,.
\end{align}
Using the fact that, to zeroth-order in $\kappa$, the energy-momentum tensor is $T_{\mu\nu}\approx-2\delta S_{\text m}/\delta g^{\mu\nu}|_{g=\eta}$, Eq.~\eqref{eq:approx_matter_action} reduces to Eq.~\eqref{eq:matter_action}. We now wish to consider the equations of motion $\delta S_{\text G}/\delta h^{\mu\nu}=-\delta S_{\text m}/\delta h^{\mu\nu}$ for the IDG theory. By making use of Eq.~\eqref{eq:eom_h} and imposing the de Donder gauge $\partial_\nu h^\nu\phantom{}_\mu=\partial_\mu h/2$ one finds
\begin{align}\label{eq:eom_no_L}
\mathcal{F}(\Box)\Box\left(h_{\mu\nu}-\frac12\eta_{\mu\nu}h\right)=-\kappa T_{\mu\nu}\,.
\end{align}
The linearised IDG equations of motion in Eq.~\eqref{eq:eom_no_L} can be written as
\begin{align}\label{eq:eom_L}
L_{\mu\nu}\phantom{}^{\alpha\beta}\mathcal{F}(\Box)\Box h_{\alpha\beta}=-\kappa I_{\mu\nu}\phantom{}^{\alpha\beta}T_{\alpha\beta}\,,
\end{align}
where \cite{Donoghue:2017pgk}
\begin{align}
I_{\mu\nu\alpha\beta}&=\frac12\left(\eta_{\mu\alpha}\eta_{\nu\beta}+\eta_{\mu\beta}\eta_{\nu\alpha}\right)\,,\label{eq:I_tensor}\\
L_{\mu\nu\alpha\beta}&=I_{\mu\nu\alpha\beta}-\frac12\eta_{\mu\nu}\eta_{\alpha\beta}\,.\label{eq:L_tensor}
\end{align}
We note that the above expressions satisfy $L_{\mu\nu}\phantom{}^{\alpha\beta}L_{\alpha\beta}\phantom{}^{\sigma\rho}=I_{\mu\nu}\phantom{}^{\sigma\rho}$ and $L_{\mu\nu}\phantom{}^{\alpha\beta}I_{\alpha\beta}\phantom{}^{\sigma\rho}=L_{\mu\nu}\phantom{}^{\sigma\rho}$. The Green's function associated with Eq.~\eqref{eq:eom_L} satisfies
\begin{align}\label{eq:greens_satisfies}
L_{\mu\nu}\phantom{}^{\alpha\beta}\mathcal{F}(\Box)\Box G_{\alpha\beta\sigma\rho}(x,x')=I_{\mu\nu\sigma\rho}\delta^{(4)}(x-x')\,.
\end{align}
By acting $L_{\alpha\beta}\phantom{}^{\mu\nu}$ on Eq.~\eqref{eq:greens_satisfies} one obtains
\begin{align}\label{eq:greens_satisfies_projected}
\mathcal{F}(\Box)\Box G_{\mu\nu\alpha\beta}(x,x')=L_{\mu\nu\alpha\beta}\delta^{(4)}(x-x')\,.
\end{align}
Finally, taking the Fourier transform of Eq.~\eqref{eq:greens_satisfies_projected}, dividing both sides by $-k^2\mathcal{F}(-k^2)$ and then taking the inverse Fourier transform yields Eq.~\eqref{eq:Green's_function_linearised_IDG}.
\section{Derivations of Eqs.~\eqref{eq:T_mu_nu_massive_kets}, \eqref{eq:maxwell_field_energy_momentum_tensor_final} and \eqref{eq:spinor_field_energy_momentum_tensor_final}}\label{sec:quantum_energy_momentum_tensors_derivations}
In this section, we review the derivations of the quantum energy-momentum tensor expressions for the spinless, Maxwell and Dirac fields. While in the present work we make use of a mostly positive signature, such expressions for a mostly negative signature can be found in~\cite{michael1979advanced}.
\subsection{Derivation of Eq.~\eqref{eq:T_mu_nu_massive_kets}}
The action for a massive scalar field is given by
\begin{align}
S_{\text m}=-\frac12\int\text{d}^4x\sqrt{-g}\left[\left(\nabla\phi\right)^2+m^2\phi^2\right]\,,
\end{align}
where $\left(\nabla\phi\right)^2=g^{\mu\nu}\nabla_\mu\phi\nabla_\nu\phi$. Expanded up to zeroth-order in $\kappa$, the action is
\begin{align}\label{eq:linearised_action}
S_{\text m}\approx-\frac12\int\text{d}^4x\left[\left(\partial\phi\right)^2+m^2\phi^2\right]\,,
\end{align}
where $\left(\partial\phi\right)^2=\eta^{\mu\nu}\partial_\mu\phi\partial_\nu\phi$. The energy-momentum tensor may be obtained via Eq.~\eqref{eq:energy_momentum_def}. In what follows, we shall write the energy-momentum tensor in terms of creation and annihilation operators after performing the mode expansion for the scalar field and shall therefore write it using normal ordering. To zeroth-order in $\kappa$, it is given by
\begin{align}\label{eq:linearised_em}
T_{\mu\nu}=:\left\{\partial_\mu\phi\partial_\nu\phi-\frac12\eta_{\mu\nu}\left[\left(\partial\phi\right)^2+m^2\phi^2\right]\right\}:\,,
\end{align}
where the $::$ indicate normal ordering. In the linearised regime, the equation of motion for the scalar field is
\begin{align}
\left(\Box-m^2\right)\phi=0\,,
\end{align}
where $\Box=\eta^{\mu\nu}\partial_\mu\partial_\nu$ as before and we can perform the following Fourier expansion of the scalar field~\cite{Peskin:1995ev}
\begin{align}
\phi(x)=\frac{1}{(2\pi)^3}\int\frac{\text{d}^3k}{\sqrt{2\omega_{\bm k}}}\left[a^\dagger(\bm k){\text e}^{-ik\cdot x}+a(\bm k){\text e}^{ik\cdot x}\right]\,,
\end{align}
where $\omega_{\bm k}=k^0=\sqrt{\bm k^2+m^2}$. The canonical conjugate momentum for $\phi$ is $\Pi=\partial\mathcal{L}/\partial\dot\phi=\dot\phi$ where $\mathcal L$ is the Lagrangian density in the linearised action~\eqref{eq:linearised_action}. The standard equal time commutation relation is
\begin{align}\label{eq:comm_phi_Pi}
\left[\phi(t,\bm x),\Pi(t,\bm x')\right]=i\delta^{(3)}(\bm x-\bm x')\,,
\end{align}
while the commutation relation for the creation and annihilation operators is
\begin{align}\label{eq:comm_phi}
\left[a(\bm k),a^\dagger(\bm k')\right]=(2\pi)^3\delta^{(3)}(\bm k-\bm k')\,.
\end{align}
It is not difficult to verify that the commutation relation~\eqref{eq:comm_phi} leads to~\eqref{eq:comm_phi_Pi} being satisfied. We also note that
\begin{align}
\partial_\mu\phi=-\frac{i}{(2\pi)^3}\int\frac{\text{d}^3k}{\sqrt{2\omega_{\bm k}}}k_\mu\left[a^\dagger(\bm k){\text e}^{-ik\cdot x}-a(\bm k){\text e}^{ik\cdot x}\right]\,,
\end{align}
from which one can compute
\begin{align}
\partial_\mu\phi\partial_\nu\phi=-\frac1{(2\pi)^6}\int\frac{\text{d}^3k\text{d}^3k'}{\sqrt{4\omega_{\bm k}\omega_{\bm k'}}}k_\mu k_\nu'\bigg[a^\dagger(\bm k)a^\dagger(\bm k'){\text e}^{-ix\cdot(k+k')}-a(\bm k)a^\dagger(\bm k'){\text e}^{ix\cdot(k-k')}\nonumber\\
-a^\dagger(\bm k)a(\bm k'){\text e}^{ix\cdot(k'-k)}+a(\bm k)a(\bm k'){\text e}^{ix\cdot(k+k')}\bigg]\,.
\end{align}
We define the relativistically normalised state $\ket p=\sqrt{2\omega_{\bm p}}a^\dagger(\bm p)\ket0$ where $\ket0$ is the vacuum state and satisfies the normalisation condition $\braket{0|0}=1$. When writing the energy-momentum tensor in terms of creation and annihilation operators, we take the normal ordered product of said operators. To this end, we compute $\bra{p'}:\partial_\mu\phi\partial_\nu\phi:\ket p$ as follows
\begin{align}
\bra{p'}:\partial_\mu\phi\partial_\nu\phi:\ket p=\left(2\pi\right)^{-6}\bra0\int&\frac{\text{d}^3k\text{d}^3k'}{\sqrt{4\omega_{\bm k}\omega_{\bm k'}}}\sqrt{4\omega_{\bm p}\omega_{\bm p'}}k_\mu k'_\nu\bigg[a(\bm p')a^\dagger(\bm k')a(\bm k)a^\dagger(\bm p){\text e}^{ix\cdot(k-k')}\nonumber\\
&+a(\bm p')a^\dagger(\bm k)a(\bm k')a^\dagger(\bm p){\text e}^{ix\cdot(k'-k)}\bigg]\ket0\,.
\end{align}
Making use of the commutation relation~\eqref{eq:comm_phi} yields
\begin{align}
\bra{p'}:\partial_\mu\phi\partial_\nu\phi:\ket p=\bra0\int&\frac{\text{d}^3k\text{d}^3k'}{\sqrt{4\omega_{\bm k}\omega_{\bm k'}}}\sqrt{4\omega_{\bm p}\omega_{\bm p'}}k_\mu k'_\nu\bigg[\delta^{(3)}(\bm p'-\bm k')\delta^{(3)}(\bm k-\bm p){\text e}^{ix\cdot(k-k')}\nonumber\\
&+\delta^{(3)}(\bm p'-\bm k)\delta^{(3)}(\bm k'-\bm p){\text e}^{ix\cdot(k'-k)}\bigg]\ket0\,,
\end{align}
and thus, upon the evaluation of the integrals,
\begin{align}\label{eq:inner_product_partial_phi}
\bra{p'}:\partial_\mu\phi\partial_\nu\phi:\ket p={\text e}^{ix\cdot(p-p')}\left(p_\mu p'_\nu+p_\nu p'_\mu\right)\,.
\end{align}
Taking the trace of Eq.~\eqref{eq:inner_product_partial_phi} gives $\bra{p'}:(\partial\phi)^2:\ket p=2{\text e}^{ix\cdot(p-p')}p'\cdot p$ and it is not difficult to verify that $\bra{p'}:\phi^2:\ket p=2{\text e}^{ix\cdot(p-p')}$. Using these expressions together with Eq.~\eqref{eq:linearised_em} gives Eq.~\eqref{eq:T_mu_nu_massive_kets}.
\subsection{Derivation of Eq.~\eqref{eq:maxwell_field_energy_momentum_tensor_final}}
The action for the Maxwell field is given by
\begin{align}\label{eq:full_action_Maxwell}
S_{\text m}=-\frac14\int\text{d}^4x\sqrt{-g}g^{\mu\alpha}g^{\nu\beta}F_{\mu\nu}F_{\alpha\beta}\,,
\end{align}
where the field strength tensor is $F_{\mu\nu}=2\partial_{[\mu}A_{\nu]}$ and $A^\mu$ is the four-vector potential. Expanded up to zeroth-order in $\kappa$, the action~\eqref{eq:full_action_Maxwell} for the Maxwell field is~\cite{Peskin:1995ev}
\begin{align}\label{eq:linear_action_Maxwell}
S_{\text m}\approx-\frac14\int\text{d}^4xF_{\mu\nu}F^{\mu\nu}\,.
\end{align}
Here, we work in the Lorentz gauge and add the gauge-fixing action $S_{\text{LGF}}=-\int\text{d}^4x\left(\partial_\mu A^\mu\right)^2/2$. The addition of such a gauge-fixing action to the linearised action~\eqref{eq:linear_action_Maxwell} then yields
\begin{align}\label{eq:linear_action_Maxwell_gf}
S_{\text m}+S_{\text{LGF}}\approx-\frac12\int\text{d}^4x\partial_\mu A_\nu\partial^\mu A^\nu\,,
\end{align}
whose equations of motion are
\begin{align}\label{eq:eom_A}
\Box A^\mu=0\,.
\end{align}
The corresponding energy-momentum tensor can be found by applying Eq.~\eqref{eq:energy_momentum_def} to Eq.~\eqref{eq:full_action_Maxwell} and, to zeroth-order in $\kappa$ with normal ordering, it is
\begin{align}\label{eq:T_mu_nu_F_2}
T_{\mu\nu}=:\left(F_{\mu\alpha}F_\nu\phantom{}^\alpha-\frac14\eta_{\mu\nu}F_{\alpha\beta}F^{\alpha\beta}\right):\,.
\end{align}
We expand the four-vector potential in terms of creation and annihilation operators as follows
\begin{align}
A^\mu(x)=\frac{1}{(2\pi)^3}\int\frac{\text{d}^3k}{\sqrt{2|\bm k|}}\epsilon^{\mu\alpha}(\bm k)\left[a^\dagger_\alpha(\bm k){\text e}^{-ik\cdot x}+a_\alpha(\bm k){\text e}^{ik\cdot x}\right]\,,
\end{align}
where $k^2=0$ which allows for~\eqref{eq:eom_A} to be satisfied\footnote{It is worth pointing out that, since the Lorentz gauge-fixing term has been added to Eq.~\eqref{eq:linear_action_Maxwell}, it is necessary to introduce the so-called \textit{Gupta-Bleuler supplementary condition}~\cite{Suraj_N_Gupta_1950,Bleuler:1950cy}
\begin{align}
\partial^\mu A_\mu^{(+)}\ket\psi=0\,,
\end{align}
for physical states $\ket\psi$ and where the $(+)$ superscript refers to only the ${\text e}^{+ik\cdot x}$ part in the mode expansion of $A^\mu$. Such a condition ensures that there are no negative norm states and that the Hamiltonian for the free Maxwell field is bounded from below. While we do not examine such a condition in detail here, we refer the reader  to~\cite{tong_lecture,Suraj_N_Gupta_1950,Bleuler:1950cy} for details.}. We note that the polarisation vectors satisfy $\epsilon^\alpha\phantom{}_\tau(\bm k)\epsilon_{\alpha\tau'}(\bm k)=\eta_{\tau\tau'}$ and $\epsilon_\mu\phantom{}^\tau(\bm k)\epsilon_{\nu\tau}(\bm k)=\eta_{\mu\nu}$ \cite{tong_lecture}. The canonical conjugate momentum for the four-vector potential is $\Pi_\mu=\partial\mathcal L/\partial \dot A^\mu=\dot A_\mu$ where $\mathcal L$ is the Lagrangian density in Eq.~\eqref{eq:linear_action_Maxwell_gf}. The equal time commutation relation is
\begin{align}\label{eq:commutation_Maxwell_2}
\left[A_\mu(t,\bm x),\Pi_\nu(t,\bm x')\right]=i\eta_{\mu\nu}\delta^{(3)}(\bm x-\bm x')\,,
\end{align}
and it is not difficult to show that
\begin{align}\label{eq:commutation_Maxwell}
\left[a_\mu(\bm k),a^\dagger_\nu(\bm k')\right]=\eta_{\mu\nu}(2\pi)^3\delta^{(3)}(\bm k-\bm k')\,,
\end{align}
leads to Eq.~\eqref{eq:commutation_Maxwell_2} being satisfied.

In order to derive Eq.~\eqref{eq:maxwell_field_energy_momentum_tensor_final}, we first compute
\begin{align}
\partial_\mu A^\alpha\partial_\nu A_\alpha=-\frac1{(2\pi)^6}\int\frac{\text{d}^3k\text{d}^3k'}{\sqrt{4|\bm k||\bm k'|}}k_\mu k'_\nu\epsilon^{\alpha\beta}(\bm k)\epsilon_\alpha\phantom{}^\sigma(\bm k')\bigg[a^\dagger_\beta(\bm k)a^\dagger_\sigma(\bm k'){\text e}^{-ix\cdot(k+k')}\nonumber\\
-a_\beta(\bm k)a^\dagger_\sigma(\bm k'){\text e}^{ix\cdot(k-k')}-a_\beta^\dagger(\bm k)a_\sigma(\bm k'){\text e}^{ix\cdot(k'-k)}+a_\beta(\bm k)a_\sigma(\bm k'){\text e}^{ix\cdot(k+k')}\bigg]\,.
\end{align}
By defining the states $\ket{p,\tau}=\sqrt{2|\bm p|}a^\dagger_\tau(\bm p)\ket0$, one finds
\begin{align}
&\bra{\tau',p'}:\partial_\mu A^\alpha\partial_\nu A_\alpha:\ket{p,\tau}=(2\pi)^{-6}\bra0\int\frac{\text{d}^3k\text{d}^3k'}{\sqrt{4|\bm k||\bm k'|}}\sqrt{4|\bm p||\bm p'|}\epsilon^{\alpha\beta}(\bm k)\epsilon_\alpha\phantom{}^\sigma(\bm k')k_\mu k'_\nu\nonumber\\
&\times\bigg[a_{\tau'}(\bm p')a^\dagger_\sigma(\bm k')a_\beta(\bm k)a^\dagger_\tau(\bm p){\text e}^{ix\cdot(k-k')}+a_{\tau'}(\bm p')a^\dagger_\beta(\bm k)a_\sigma(\bm k')a^\dagger_\tau(\bm p){\text e}^{ix\cdot(k'-k)}\bigg]\ket0\,.
\end{align}
Using the commutation relation~\eqref{eq:commutation_Maxwell} leads to
\begin{align}
\bra{\tau',p'}:\partial_\mu A^\alpha\partial_\nu A_\alpha:\ket{p,\tau}=\epsilon^\alpha\phantom{}_\tau(\bm p)\epsilon^\beta\phantom{}_{\tau'}(\bm p'){\text e}^{ix\cdot(p-p')}\eta_{\alpha\beta}\left(p_\mu p'_\nu+p_\nu p'_\mu\right)\,.
\end{align}
Similarly, one can obtain
\begin{align}
\bra{\tau',p'}:\partial_\alpha A_\mu\partial_\nu A^\alpha:\ket{p,\tau}&=\epsilon^\alpha\phantom{}_\tau(\bm p)\epsilon^\beta\phantom{}_{\tau'}(\bm p'){\text e}^{ix\cdot(p-p')}\left(p'_\alpha p_\nu\eta_{\beta\mu}+p_\beta p'_\nu\eta_{\alpha\mu}\right)\,,\\
\bra{\tau',p'}:\partial_\alpha A_\mu\partial^\alpha A_\nu:\ket{p,\tau}&=\epsilon^\alpha\phantom{}_\tau(\bm p)\epsilon^\beta\phantom{}_{\tau'}(\bm p'){\text e}^{ix\cdot(p-p')}p\cdot p'\left(\eta_{\alpha\mu}\eta_{\beta\nu}+\eta_{\alpha\nu}\eta_{\beta\mu}\right)\,.
\end{align}
One now has
\begin{align}\label{eq:f_2_1}
\bra{\tau',p'}:F_{\mu\alpha}F_\nu\phantom{}^\alpha:\ket{p,\tau}=&\epsilon^\alpha\phantom{}_\tau(\bm p)\epsilon^\beta\phantom{}_{\tau'}(\bm p'){\text e}^{ix\cdot(p-p')}\bigg[\eta_{\alpha\beta}\left(p_\mu p'_\nu+p'_\mu p_\nu\right)+p\cdot p'\left(\eta_{\mu\alpha}\eta_{\nu\beta}+\eta_{\nu\alpha}\eta_{\mu\beta}\right)\nonumber\\
&-p_\alpha'p_\nu\eta_{\beta\mu}-p_\beta p'_\nu\eta_{\alpha\mu}-p'_\alpha p_\mu\eta_{\beta\nu}-p_\beta p'_\mu\eta_{\alpha\nu}\bigg]\,,
\end{align}
as well as its trace
\begin{align}\label{eq:f_2_2}
\bra{\tau',p'}:F_{\mu\nu}F^{\mu\nu}:\ket{p,\tau}=4\epsilon^\alpha\phantom{}_\tau(\bm p)\epsilon^\beta\phantom{}_{\tau'}(\bm p'){\text e}^{ix\cdot(p-p')}\left(p\cdot p'\eta_{\alpha\beta}-p'_\alpha p_\beta\right)\,.
\end{align}
Substituting Eqs.~\eqref{eq:f_2_1} and~\eqref{eq:f_2_2} into Eq.~\eqref{eq:T_mu_nu_F_2} yields Eq.~\eqref{eq:maxwell_field_energy_momentum_tensor_final}.
\subsection{Derivation of Eq.~\eqref{eq:spinor_field_energy_momentum_tensor_final}}
The linearised action for the Dirac field is given by~\cite{Bandyopadhyay}
\begin{align}\label{eq:dirac_matter_action}
S_{\text m}=\frac12\int\text{d}^4x\bar\psi\left(i\gamma^\mu\overrightarrow{\partial}_\mu-i\gamma^\mu\overleftarrow{\partial}_\mu-2m\right)\psi\,.
\end{align}
By computing $-\left(\frac{\partial\mathcal L}{\partial\partial_\mu\psi}\partial_\nu\psi+\partial_\nu\bar\psi\frac{\partial\mathcal L}{\partial\partial_\mu\bar\psi}-\mathcal L\right)$, where $\mathcal L$ is the Lagrangian density in Eq.~\eqref{eq:dirac_matter_action}, symmetrising the result and using normal ordering, one obtains\footnote{For details regarding the full construction of the symmetric energy-momentum tensor in Eq.~\eqref{eq:t_mu_nu_dirac}, we direct the reader to~\cite{Bandyopadhyay}.}
\begin{align}\label{eq:t_mu_nu_dirac}
T_{\mu\nu}=-\frac i4:\bar\psi\left(\gamma_\mu\overrightarrow{\partial}_\nu+\gamma_\nu\overrightarrow{\partial}_\mu-\gamma_\mu\overleftarrow{\partial}_\nu-\gamma_\nu\overleftarrow{\partial}_\mu\right)\psi:\,.
\end{align}
The Dirac fields are expanded in terms of creation and annihilation operators as follows~\cite{Peskin:1995ev}
\begin{align}
\psi(x)&=\frac{1}{(2\pi)^3}\sum_s\int\frac{\text{d}^3k}{\sqrt{2\omega_{\bm k}}}\left[b^\dagger_s(\bm k)v_s(k){\text e}^{-ik\cdot x}+c_s(\bm k)u_s(k){\text e}^{ik\cdot x}\right]\,,\\
\bar\psi(x)&=\frac{1}{(2\pi)^3}\sum_s\int\frac{\text{d}^3k}{\sqrt{2\omega_{\bm k}}}\left[c^\dagger_s(\bm k)\bar u_s(k){\text e}^{-ik\cdot x}+b_s(\bm k)\bar v_s(k){\text e}^{ik\cdot x}\right]\,,
\end{align}
where $k^0=\omega_{\bm k}=\sqrt{\bm k^2+m^2}$. In addition, it is also noted that $u_s(p)$ is given by Eq.~\eqref{eq:u_s_def} while $v_s(p)$ is given by~\cite{Peskin:1995ev}
\begin{align}\label{eq:v_s_def}
v_s(p)=\begin{pmatrix}\sqrt{p^0-\bm\sigma\cdot\bm p}\chi_s\\ -\sqrt{p^0+\bm\sigma\cdot\bm p}\chi_s\end{pmatrix}\,,
\end{align}
where the $\chi_s$ are two-component numerical spinors which satisfy $\chi^\dagger_{s'}\chi_s=\delta_{s's}$. The anticommutation relations
\begin{align}
\left\{c_s(\bm k),c^\dagger_{s'}(\bm k')\right\}=\left\{b_s(\bm k),b^\dagger_{s'}(\bm k')\right\}=(2\pi)^3\delta_{ss'}\delta^{(3)}(\bm k-\bm k')\,,
\end{align}
lead to the equal time anticommutation relation $\left\{\psi(t,\bm x),\psi^\dagger(t,\bm x')\right\}=\delta^{(3)}(\bm x-\bm x')$ being satisfied.

In what follows, let us define the following one-particle excitation state $\ket{p,s}=\sqrt{2\omega_{\bm p}}c_s^\dagger(\bm p)\ket0$. We now compute
\begin{align}
&\bar\psi\gamma_\mu\overrightarrow\partial_\nu\psi=-\frac{i}{(2\pi)^6}\sum_{s,s'}\int\frac{\text{d}^3k\text{d}^3k'k'_\nu}{\sqrt{4\omega_{\bm k}\omega_{\bm k'}}}\bigg[c_s^\dagger(\bm k)b_{s'}^\dagger(\bm k')\bar u_s(k)\gamma_\mu v_{s'}(k'){\text e}^{-ix\cdot (k+k')}-c_s^\dagger(\bm k)c_{s'}(\bm k')\bar u_s(k)\gamma_\mu\nonumber\\
&\times u_{s'}(k'){\text e}^{ix\cdot (k'-k)}+b_s(\bm k)b_{s'}^\dagger(\bm k')\bar v_s(k)\gamma_\mu v_{s'}(k'){\text e}^{ix\cdot (k-k')}-b_s(\bm k)c_{s'}(\bm k')\bar v_s(k)\gamma_\mu u_{s'}(k'){\text e}^{ix\cdot (k+k')}\bigg]\,.\label{eq:psi_partial_psi}
\end{align}
When taking the inner product of Eq.~\eqref{eq:psi_partial_psi} with norml ordering between $\ket{p,s}$ and $\ket{p',s'}$, only the second term gives a non-vanishing contribution. We now have
\begin{align}\label{eq:psi_deriv_normal_order}
\bra{s',p'}:\bar\psi\gamma_\mu\overrightarrow\partial_\nu\psi:\ket{p,s}=i\bar u_{s'}(p')\gamma_\mu p_\nu u_s(p){\text e}^{ix\cdot(p-p')}\,,
\end{align}
and similarly
\begin{align}\label{eq:bar_psi_deriv_normal_order}
\bra{s',p'}:\bar\psi\gamma_\mu\overleftarrow{\partial}_\nu\psi:\ket{p,s}=-i\bar u_{s'}(p')\gamma_\mu p'_\nu u_s(p){\text e}^{ix\cdot(p-p')}\,.
\end{align}
Eqs.~\eqref{eq:psi_deriv_normal_order}, \eqref{eq:bar_psi_deriv_normal_order} and \eqref{eq:t_mu_nu_dirac} together give Eq.~\eqref{eq:spinor_field_energy_momentum_tensor_final}.
\section{Derivation of the $S$-matrix element Eq.~\eqref{eq:S_matrix_element_3}}\label{sec:s_matrix_derivation}
Let us start by considering the total action given in Eq.~\eqref{eq:gauged_fixed_action} which is
\begin{align}\label{eq:total_action_explicit}
S_{\text{total}}=\frac14\int\text{d}^4x\left[h^{\mu\nu}\mathcal{F}(\Box)\Box\left(h_{\mu\nu}-\frac12\eta_{\mu\nu}h\right)+2\kappa h^{\mu\nu}T_{\mu\nu}\right]\,,
\end{align}
after substituting in Eqs.~\eqref{eq:quadratic_IDG_action}, \eqref{eq:matter_action} and \eqref{eq_gauge_fixing_action}, integrating by parts and simplifying. By making use of Eqs.~\eqref{eq:I_tensor} and~\eqref{eq:L_tensor}, we can write Eq.~\eqref{eq:total_action_explicit} given above as
\begin{align}\label{eq:total_action_explicit_L}
S_{\text{total}}=\frac14\int\text{d}^4x\left[h^{\mu\nu}L_{\mu\nu\alpha\beta}\mathcal{F}(\Box)\Box h^{\alpha\beta}+2\kappa h^{\mu\nu}I_{\mu\nu\alpha\beta}T^{\alpha\beta}\right]\,,
\end{align}
and it is straightforward to verify that its equations of motion are given by Eq.~\eqref{eq:eom_L} which are the linearised IDG field equations in the de Donder gauge. The integrand in Eq.~\eqref{eq:total_action_explicit_L} can be factorised in order to obtain
\begin{align}\label{eq:final_s_total_redef}
S_{\text{total}}=\frac14\int\text{d}^4x\left[\tilde h^{\mu\nu}L_{\mu\nu\alpha\beta}\mathcal{F}(\Box)\Box\tilde h^{\alpha\beta}-\kappa^2\int\text{d}^4x'T^{\mu\nu}(x)G_{\mu\nu\alpha\beta}(x,x')T^{\alpha\beta}(x')\right]\,,
\end{align}
where we define
\begin{align}
\tilde h_{\mu\nu}=h_{\mu\nu}+\kappa\int\text{d}^4x'G_{\mu\nu\alpha\beta}(x,x')T^{\alpha\beta}(x')\,,
\end{align}
and we have made use of $L_{\mu\nu}\phantom{}^{\alpha\beta}L_{\alpha\beta}\phantom{}^{\sigma\rho}=I_{\mu\nu}\phantom{}^{\sigma\rho}$ and $L_{\mu\nu}\phantom{}^{\alpha\beta}I_{\alpha\beta}\phantom{}^{\sigma\rho}=L_{\mu\nu}\phantom{}^{\sigma\rho}$.

In order to consider the $S$-matrix element, we first examine the partition function $\mathcal{Z}=\int\mathcal{D}\bm h{\text e}^{iS_{\text{total}}}/\mathcal{N}$ where $\mathcal{N}$ normalises the partition function to unity when $T_{\mu\nu}=0$, i.e., $\mathcal{N}=\int\mathcal{D}\bm h{\text e}^{iS_{\text G}+iS_{\text{GF}}}$. Using Eq.~\eqref{eq:final_s_total_redef} in the partition function, noting that $\mathcal D\tilde{\bm h}=\mathcal D\bm h$ and then relabelling $\tilde h_{\mu\nu}$ as $h_{\mu\nu}$ yields
\begin{align}\label{eq:partition_function_final}
\mathcal{Z}={\text e}^{-\frac{i\kappa^2}4\int\text{d}^4x\text{d}^4x'T^{\mu\nu}(x)G_{\mu\nu\alpha\beta}(x,x')T^{\alpha\beta}(x')}\,.
\end{align}
Taking the inner product of Eq.~\eqref{eq:partition_function_final} between initial and final states $\ket i$ and $\ket f$, respectively, which each contain two-particle excitations yields the following $S$-matrix element
\begin{align}\label{eq:s_matrix_appendix_final}
\bra fS\ket i=\braket{f|i}-\frac{i\kappa^2}4\int\text{d}^4x\text{d}^4x'\bra fT^{\mu\nu}(x)G_{\mu\nu\alpha\beta}(x,x')T^{\alpha\beta}(x')\ket i\,.
\end{align}
At this point let us write $T^{\mu\nu}=T_A^{\mu\nu}+T_B^{\mu\nu}$, $\ket i=\ket{i_A,i_B}$ and $\ket f=\ket{f_A,f_B}$ where the $\ket{i_{A,B}}$ and $\ket{f_{A,B}}$ consist of one-particle excitations associated with the corresponding field. For such a consideration, Eq.~\eqref{eq:s_matrix_appendix_final} reduces to Eq.~\eqref{eq:S_matrix_element_3}.
\section{Derivations of scattering amplitudes}\label{sec:scattering_amplitude_derivations}
In this section, we discuss the derivations for the scattering amplitudes~\eqref{eq:amplitude_spinless}, \eqref{eq:scattering_amplitude_photon} and~\eqref{eq:sum_scattering_amplitude_spin}. Let us start by substituting the Green's function~\eqref{eq:Green's_function_linearised_IDG} into Eq.~\eqref{eq:S_matrix_element_3} for the $S$-matrix element in order to write
\begin{align}\label{eq:S_matrix_element_explicit_Greens}
\bra fS-1\ket i=\frac{i\kappa^2}{2(2\pi)^4}\int\text{d}^4x&\text{d}^4x'\text{d}^4k\frac{{\text e}^{ik\cdot(x-x')}}{\mathcal{F}(-k^2)k^2}\bigg[\bra{f_A}T_A^{\mu\nu}(x)\ket{i_A}\bra{f_B}T_{B\mu\nu}\left(x'\right)\ket{i_B}\nonumber\\
&-\frac12\bra{f_A}T_A(x)\ket{i_A}\bra{f_B}T_B\left(x'\right)\ket{i_B}\bigg]\,.
\end{align}
We now wish to separate the energy-momentum tensors into products of its position and momentum dependence before evaluating the integrals contained in~\eqref{eq:S_matrix_element_explicit_Greens}. To this end, we note that Eqs.~\eqref{eq:T_mu_nu_massive_kets}, \eqref{eq:maxwell_field_energy_momentum_tensor_final} and \eqref{eq:spinor_field_energy_momentum_tensor_final} allow us to write
\begin{align}
\bra{f_A}T_{A\mu\nu}(x)&\ket{i_A}={\text e}^{ix\cdot (p-p')}V_{A\mu\nu}(p,p')\,,\label{eq:V_A_def}\\
\bra{f_B}T_{B\mu\nu}(x)&\ket{i_B}={\text e}^{ix\cdot (q-q')}V_{B\mu\nu}(q,q')\,,
\end{align}
where the $V_A$ and $V_B$ are determined by the right-hand sides of the aforementioned expressions. In addition, the $p$ and $p'$ are, respectively, the incoming and outgoing momenta of the $A$ particle while $q$ and $q'$ are those of the $B$ particle. This allows us to write Eq.~\eqref{eq:S_matrix_element_explicit_Greens} as
\begin{align}\label{eq:S_matrix_element_explicit_Greens_V}
\bra fS-1\ket i=\frac{i\kappa^2}{2(2\pi)^4}\int\text{d}^4x&\text{d}^4x'\text{d}^4k\frac{{\text e}^{ix\cdot(p-p'+k)+ix'\cdot(q-q'-k)}}{\mathcal{F}(-k^2)k^2}\nonumber\\
&\times\left[V_A^{\mu\nu}(p,p')V_{B\mu\nu}(q,q')-\frac12V_A(p,p')V_B(q,q')\right]\,.
\end{align}
Performing the integrals over $x$ and $x'$ above gives us
\begin{align}
\bra fS-1\ket i=\frac{i\kappa^2}{2}(2\pi)^4\int\text{d}^4k&\delta^{(4)}(p-p'+k)\delta^{(4)}(q-q'-k)\frac1{\mathcal{F}(-k^2)k^2}\nonumber\\
&\times\left[V_A^{\mu\nu}(p,p')V_{B\mu\nu}(q,q')-\frac12V_A(p,p')V_B(q,q')\right]\,.
\end{align}
Now, integrating over $k$ then gives
\begin{align}\label{eq:S_matrix_V_A_B}
\bra fS-1\ket i=\frac{i\kappa^2}{2}(2\pi)^4&\delta^{(4)}(p-p'+q-q')\frac1{\mathcal{F}(-t)t}\left[V_A^{\mu\nu}(p,p')V_{B\mu\nu}(q,q')-\frac12V_A(p,p')V_B(q,q')\right]\,,
\end{align}
where we have used the Mandelstam variable $t=(p'-p)^2$. As the result of the Dirac delta function, we have $p'-p=q-q'=k$ where $k$ now denotes the graviton transfer momentum. As already mentioned, we take the $A$ and $B$ particles to be on-shell, i.e., $p^2=p'^2=-m_A^2$ and $q^2=q'^2=-m_B^2$. The on-shell condition for the $A$ particle gives
\begin{align}\label{eq:p_k_to_k}
-m_A^2=p'^2=(p+k)^2=k^2+p^2+2k\cdot p\implies2k\cdot p=-k^2\,,
\end{align}
while the on-shell condition for the $B$ particle gives
\begin{align}\label{eq:q_k_to_k}
-m_B^2=q'^2=(q-k)^2=k^2+q^2-2k\cdot q\implies2k\cdot q=k^2\,.
\end{align}
In \ref{sec:appendix_spinless}, \ref{sec:appendix_photon} and~\ref{sec:appendix_spin} we shall derive the scattering amplitudes for the single graviton exchange between two spinless particles, a spinless particle and a photon, and between a spinless particle and a spin-$1/2$ particle, respectively. In the aforementioned sections, we shall make use of Eqs.~\eqref{eq:p_k_to_k} and \eqref{eq:q_k_to_k}.
\subsection{Scattering amplitude for the case of two spinless particles}\label{sec:appendix_spinless}
For this case, the $V_{A\mu\nu}$ is given by the square brackets in Eq.~\eqref{eq:T_mu_nu_massive_kets} with $m$ replaced by $m_A$. Similarly, $V_{B\mu\nu}$ is also given by the square brackets in Eq.~\eqref{eq:T_mu_nu_massive_kets} with $p$, $p'$ and $m$ therein replaced by $q$, $q'$ and $m_B$, respectively. By substituting in $p'=p+k$ and $q'=q-k$, one arrives at the following 
\begin{align}
V_{A\mu\nu}&=2p_\mu p_\nu+p_\mu k_\nu+k_\mu p_\nu+\eta_{\mu\nu}\frac{k^2}2\,,\label{eq:V_A_spinless}\\
V_{B\mu\nu}&=2q_\mu q_\nu-q_\mu k_\nu-k_\mu q_\nu+\eta_{\mu\nu}\frac{k^2}2\,,\label{eq:V_B_spinless}\\
V_A&=2p^2+k^2\,,\label{eq:V_A_spinless_trace}\\
V_B&=2q^2+k^2\,,\label{eq:V_B_spinless_trace}
\end{align}
and we have made use of Eqs.~\eqref{eq:p_k_to_k} and \eqref{eq:q_k_to_k}. We now proceed to compute
\begin{align}\label{eq:V_A_B_mu_spinless}
V_A^{\mu\nu}V_{B\mu\nu}&=4(p\cdot q)^2-4(p\cdot q)(p\cdot k)+p^2k^2+4(p\cdot q)(q\cdot k)-2(p\cdot q)k^2-2(p\cdot k)(q\cdot k)\nonumber\\
&+(p\cdot k)k^2+q^2k^2-(q\cdot k)k^2+k^4\nonumber\\
&=4(p\cdot q)^2+2(p\cdot q)k^2+p^2k^2+q^2k^2+\frac{k^4}2\,,
\end{align}
where, after the second equality, we have once again made use of Eqs.~\eqref{eq:p_k_to_k} and \eqref{eq:q_k_to_k}. We also have
\begin{align}\label{eq:V_A_B_spinless}
V_AV_B=4p^2q^2+2p^2k^2+2q^2k^2+k^4\,.
\end{align}
Thus, upon the subtraction of half Eq.~\eqref{eq:V_A_B_spinless} from Eq.~\eqref{eq:V_A_B_mu_spinless} and the substitution of the result into Eq.~\eqref{eq:S_matrix_V_A_B}, one finds
\begin{align}
\braket{f|S-1|i}=i\kappa^2(2\pi)^4\delta^{(4)}(p-p'+q-q')\frac1{\mathcal{F}(-t)t}\left[2\left(p\cdot q\right)^2-p^2q^2+tp\cdot q\right]\,,
\end{align}
where $t=k^2$ as before. Extracting the scattering amplitude from this $S$-matrix element through Eq.~\eqref{eq:extract_scattering_amplitude} yields Eq.~\eqref{eq:amplitude_spinless}.
\setcounter{footnote}{0}

Let us now derive the gravitational potential. Here, we work in the CM frame which has $\bm p=-\bm q$ and implies that $\bm p'=-\bm q'$, $p'^0=p^0$ and $q'^0=q^0$\footnote{Since $\bm p+\bm q=\bm p'+\bm q'$, taking $\bm p=-\bm q$ results in $\bm p'=-\bm q'$. As the result of energy conservation, we also have $p'^0=p^0+q^0-q'^0$ and squaring both sides gives
\begin{align}\label{eq:cm_frame_energy_derive_0}
m_A^2+\bm p'^2=m_A^2+\bm p^2+m_B^2+\bm q^2+m_B^2+\bm q'^2+2p^0q^0-2p^0q'^0-2q^0q'^0\,,
\end{align}
where $m_A$ and $m_B$ are the masses of the $A$ and $B$ particles, respectively. After using  $\bm q=-\bm p$ and $\bm q'=-\bm p'$, Eq.~\eqref{eq:cm_frame_energy_derive_0} gives
\begin{align}\label{eq:cm_frame_energy_derive}
m_B^2+\bm p^2+p^0q^0-p^0q'^0-q^0q'^0=0\,.
\end{align}
Using the fact that $\left(q^0\right)^2=m_B^2+\bm p^2$ in the CM frame, Eq.~\eqref{eq:cm_frame_energy_derive} becomes $q^0(q^0+p^0)=q'^0(p^0+q^0)$ and thus $q^0=q'^0$ and it follows that $p^0=p'^0$.}. For the four-momenta of the two particles, we write $p=(E_p,\bm p)$ and $q=(E_q,-\bm p)$ and therefore
\begin{align}
(p\cdot q)^2&=(E_pE_q+\bm p^2)^2=E_p^2E_q^2+2E_pE_q\bm p^2+\bm p^4\,,\label{eq:p_dot_q_squared}\\
p^2q^2&=(-E_p^2+\bm p^2)(-E_q^2+\bm p^2)=E_p^2E_q^2-\left(E_p^2+E_q^2\right)\bm p^2+\bm p^4\,,\label{eq:p_q_squared}
\end{align}
and thus
\begin{align}
2(p\cdot q)^2-p^2q^2&=E_p^2E_q^2+(4E_pE_q+E_p^2+E_q^2)\bm p^2+\bm p^4\nonumber\\
&=E_p^2E_q^2\left[1+\left(4+\frac{E_p}{E_q}+\frac{E_q}{E_p}\right)\frac{\bm p^2}{E_pE_q}+\frac{\bm p^4}{E_p^2E_q^2}\right]\,.
\end{align}
We also note that, in the CM frame, we have $t=\bm k^2$. The scattering amplitude is now
\begin{align}
\mathcal{A}(\bm k)=\frac{16\pi G}{\mathcal{F}(-\bm k^2)\bm k^2}E_p^2E_q^2\left[1+\left(4+\frac{E_p}{E_q}+\frac{E_q}{E_p}\right)\frac{\bm p^2}{E_pE_q}+\frac{\bm p^4}{E_p^2E_q^2}\right]-\frac{16\pi G}{\mathcal{F}(-\bm k^2)}E_pE_q\left(1+\frac{\bm p^2}{E_pE_q}\right)\,,
\end{align}
and we have used $\kappa^2=16\pi G$ which is true in natural units. The gravitational potential is now computed via~\cite{Barker:1966zz,michael1979advanced}
\begin{align}\label{eq:potential_from_scattering_amplitude}
V(\bm r)=-\frac{1}{4E_pE_q}\int\frac{\text{d}^3k}{(2\pi)^3}{\text e}^{i\bm k\cdot\bm r}\mathcal{A}(\bm k)\,,
\end{align}
from which one can obtain Eq.~\eqref{eq:spinless_potential_general} after introducing $c$ and $\hbar$ through dimensional analysis.
\subsection{Scattering amplitude for the case of a photon and a spinless particle}\label{sec:appendix_photon}
The initial and final states for the photon are $\ket{i_A}=\ket{p,s}=\sqrt{2|\bm p|}a_s^\dagger(\bm p)\ket0$ and $\ket{f_A}=\ket{p',s'}=\sqrt{2|\bm p'|}a_{s'}^\dagger(\bm p')\ket0$, respectively, while those of the spinless particle are the same as the previous section, i.e., $\ket{i_B}=\ket q=\sqrt{2\omega_{\bm q}}a^\dagger(\bm q)\ket0$ and $\ket{f_B}=\ket{q'}=\sqrt{2\omega_{\bm q'}}a^\dagger(\bm q')\ket0$. From Eq.~\eqref{eq:maxwell_field_energy_momentum_tensor_final} together with Eq.~\eqref{eq:V_A_def} we have
\begin{align}\label{eq:V_A_photon}
&V^{(s,s')}_{A\mu\nu}=\epsilon^i\phantom{}_s(\bm p)\epsilon^j\phantom{}_{s'}(\bm p')\bigg[\delta_{ij}\left(2p_\mu p_\nu+k_\mu p_\nu+k_\nu p_\mu\right)-\eta_{j\nu}p_\mu p_i-\eta_{j\nu}p_\mu k_i-\eta_{i\nu}p_jp_\mu-\eta_{i\nu}p_j k_\mu\nonumber\\
&-\eta_{j\mu}p_\nu p_i-\eta_{j\mu}p_\nu k_i-\eta_{i\mu}p_jp_\nu-\eta_{i\mu}p_jk_\nu-\frac{k^2}2\left(\eta_{i\mu}\eta_{j\nu}+\eta_{i\nu}\eta_{j\mu}-\delta_{ij}\eta_{\mu\nu}\right)+\eta_{\mu\nu}p_jp_i+\eta_{\mu\nu}p_jk_i\bigg]\,,
\end{align}
where we have used $p'=p+k$ and $p\cdot k=-k^2/2$ and also included the indices $s$ and $s'$ to indicate the transverse polarisation states. Since we have taken $\bm\epsilon_1$ and $\bm\epsilon_2$ to be the transverse polarisation vectors, they are orthogonal to their corresponding photon three-momentum. Therefore, $\epsilon^i\phantom{}_s(\bm p)p_i=0$ and such terms vanish in Eq.~\eqref{eq:V_A_photon} and we are left with
\begin{align}\label{eq:V_A_photon_2}
&V^{(s,s')}_{A\mu\nu}=\epsilon^i\phantom{}_s(\bm p)\epsilon^j\phantom{}_{s'}(\bm p')\bigg[\delta_{ij}\left(2p_\mu p_\nu+k_\mu p_\nu+k_\nu p_\mu\right)-\eta_{j\nu}p_\mu k_i-\eta_{i\nu}p_jp_\mu-\eta_{i\nu}p_j k_\mu\nonumber\\
&-\eta_{j\mu}p_\nu k_i-\eta_{i\mu}p_jp_\nu-\eta_{i\mu}p_jk_\nu-\frac{k^2}2\left(\eta_{i\mu}\eta_{j\nu}+\eta_{i\nu}\eta_{j\mu}-\delta_{ij}\eta_{\mu\nu}\right)+\eta_{\mu\nu}p_jk_i\bigg]\,.
\end{align}
At this point, let us impose the CM frame, i.e., $\bm p=-\bm q$ and $\bm p'=-\bm q'$ which gives $p^0=p'^0$, $q^0=q'^0$ and $k^2=\bm k^2$. As the result of working in the CM frame, we also have $\epsilon^i\phantom{}_s(\bm p)q_i=\epsilon^i\phantom{}_{s'}(\bm p')q'_i=0$. With this in mind, let us now contract Eq.~\eqref{eq:V_A_photon_2} with Eq.~\eqref{eq:V_B_spinless} and compute
\begin{align}
V^{(s,s')}_{A\mu\nu}&V_B^{\mu\nu}=\epsilon^i\phantom{}_s(\bm p)\epsilon^j\phantom{}_{s'}(\bm p')\bigg\{2\delta_{ij}\left[2\left(p\cdot q\right)^2+k^2(p\cdot q)\right]-2k_i\bigg[2\left(p\cdot q\right)q_j+\frac{k^2}{2}q_j-(p\cdot q)k_j\nonumber\\
&+\frac{k^2}2p_j\bigg]+2p_jk_i(p\cdot q)+k^2k_iq_j-\frac{k^4\delta_{ij}}{2}+\left(\frac{\delta_{ij}k^2}{2}+p_jk_i\right)\left(2q^2+k^2\right)\bigg\}\,,
\end{align}
where we have made use of $\epsilon^i\phantom{}_s(\bm p)q_i=\epsilon^i\phantom{}_s(\bm p)p_i=0$ as well as $q\cdot k=-p\cdot k=k^2/2$. Both the $k^2k_i$ terms and the $k^4$ terms cancel. In addition, since $\epsilon^j\phantom{}_{s'}(\bm p')q'_j=\epsilon^j\phantom{}_{s'}(\bm p')p'_j=0$, we can make use of $\epsilon^j\phantom{}_{s'}(\bm p')q_j=-\epsilon^j\phantom{}_{s'}(\bm p')p_j=\epsilon^j\phantom{}_{s'}(\bm p')k_j$. We now have
\begin{align}\label{eq:V_A_B_photon_final}
V^{(s,s')}_{A\mu\nu}V_B^{\mu\nu}&=\epsilon^i\phantom{}_s(\bm p)\epsilon^j\phantom{}_{s'}(\bm p')\bigg\{2\delta_{ij}\left[2\left(p\cdot q\right)^2+k^2(p\cdot q)\right]-4k_ik_j(p\cdot q)+2q^2\left(\frac{\delta_{ij}k^2}{2}-k_ik_j\right)\bigg\}\,.
\end{align}

It is straightforward to verify that the trace of Eq.~\eqref{eq:maxwell_field_energy_momentum_tensor_final} is zero and therefore the term $V_{A}^{(s,s')}V_B$ vanishes. Thus, upon the substitution of Eq.~\eqref{eq:V_A_B_photon_final} into Eq.~\eqref{eq:S_matrix_V_A_B} and keeping in mind that $t=k^2$, one arrives at the $S$-matrix element Eq.~\eqref{eq:scattering_matrix_photon}. The extraction of the scattering amplitude via Eq.~\eqref{eq:extract_scattering_amplitude} yields the following scattering amplitude
\begin{align}\label{eq:scattering_amplitude_photon_no_sum}
\mathcal{A}_{ss'}=\frac{\kappa^2\epsilon^i\phantom{}_s(\bm p)\epsilon^j\phantom{}_{s'}(\bm p')}{\mathcal{F}(-\bm k^2)\bm k^2}\bigg\{\delta_{ij}\left[2(p\cdot q)^2+\bm k^2p\cdot q+\frac{\bm k^2q^2}{2}\right]-k_ik_j\left(2p\cdot q+q^2\right)\bigg\}\,,
\end{align}
where we have used $t=k^2=\bm k^2$ in the CM frame.

We now wish to perform the average of the scattering amplitude over the transverse polarisations $s$ and $s'$. We first wish to compute the summation of $\bm\epsilon_s(\bm p)\cdot\bm\epsilon_{s'}(\bm p')$. As mentioned in the main text, we take $\bm\epsilon_1(\bm p)=\bm\epsilon_1(\bm p')=\left(\bm p\times\bm p'\right)/|\bm p\times\bm p'|$ which is orthogonal to both the incoming and outgoing photon momenta. Since these two polarisation vectors coincide, we shall simply denote them as $\bm \epsilon_1$. We then take $\bm\epsilon_2(\bm p)=\left(\bm p\times\bm \epsilon_1\right)/|\bm p|$ and $\bm \epsilon_2(\bm p')=\left(\bm p'\times\bm\epsilon_1\right)/|\bm p'|$. Clearly, $\bm\epsilon_1(\bm p)\cdot\bm\epsilon_1(\bm p')=\bm\epsilon_1^2=1$. We also have
\begin{align}
\bm p^2\bm\epsilon_2(\bm p)\cdot\bm\epsilon_2(\bm p')&=\left(\bm p\times\bm\epsilon_1\right)\cdot\left(\bm p'\times\bm\epsilon_1\right)\nonumber\\
&=\bm p\cdot\bm p'-\left(\bm p\cdot\bm\epsilon_1\right)\left(\bm p'\cdot\bm\epsilon_1\right)\,,
\end{align}
where we have used the fact that, in the CM frame, $p^0=p'^0$  which gives $|\bm p|=|\bm p'|$. The second term vanishes since $\bm\epsilon_1$ is orthogonal to both the incoming and outgoing photon momenta. In addition, using $p'=p+k$ and $p\cdot k=-k^2/2$, we have
\begin{align}
\bm\epsilon_2(\bm p)\cdot\bm\epsilon_2(\bm p')=1-\frac{\bm k^2}{2\bm p^2}\,,
\end{align}
and thus
\begin{align}\label{eq:eps_dot_eps}
\bm\epsilon_1^2+\epsilon_2(\bm p)\cdot\bm\epsilon_2(\bm p')=2-\frac{\bm k^2}{2\bm p^2}\,.
\end{align}
Dividing by a factor of $2$ yields Eq.~\eqref{eq:spin_averaged_sum_1}.

Let us now turn our attention to deriving Eq.~\eqref{eq:spin_averaged_sum_2}. In the CM frame Eq.~\eqref{eq:p_k_to_k} is $\bm k^2=2\bm p\cdot(\bm p-\bm p')$. Now, $\bm p\cdot\bm p'=|\bm p||\bm p'|\cos\theta$ where $\theta$ is the angle between the incoming and outgoing momenta of the photon. Using the fact that $|\bm p|=|\bm p'|$ then gives $\bm k^2=2\bm p^2(1-\cos\theta)$. It is then straightforward to show that
\begin{align}\label{eq:sin_theta_k}
\bm p^2\sin^2\theta=\bm k^2\left(1-\frac{\bm k^2}{4\bm p^2}\right)\,.
\end{align}
Since $\bm\epsilon_1\cdot\bm k=0$, we are left with needing to compute $\bm\epsilon_2(\bm p)\cdot\bm k$ and $\bm\epsilon_2(\bm p')\cdot\bm k$. In order to compute the former, we first note that
\begin{align}
\left[\bm p\times(\bm p\times\bm p')\right]^i=-\frac{\bm k^2p^i}2-\bm p^2k^i\,,
\end{align}
and thus
\begin{align}\label{eq:k_dot_p_cross}
\bm k\cdot\left[\bm p\times(\bm p\times\bm p')\right]=-\bm k^2\bm p^2\left(1-\frac{\bm k^2}{4\bm p^2}\right)\,.
\end{align}
It is not difficult to check that this is the same result for $\bm k\cdot\left[\bm p'\times(\bm p\times\bm p')\right]$. We can now compute the following
\begin{align}
\sum_{s,s'}\bm\epsilon_s(\bm p)\cdot\bm k\ \bm\epsilon_{s'}(\bm p')\cdot\bm k&=\frac{\left\{\bm k\cdot\left[\bm p\times(\bm p\times\bm p')\right]\right\}\left\{\bm k\cdot\left[\bm p'\times(\bm p\times\bm p')\right]\right\}}{\bm p^2|\bm p\times\bm p'|^2}\nonumber\\
&=\frac{\left\{\bm k\cdot\left[\bm p\times(\bm p\times\bm p')\right]\right\}^2}{\bm p^6\sin^2\theta}\,.
\end{align}
Substituting in Eqs.~\eqref{eq:sin_theta_k} and \eqref{eq:k_dot_p_cross} gives
\begin{align}\label{eq:eps_dot_k}
\sum_{s,s'}\bm\epsilon_s(\bm p)\cdot\bm k\ \bm\epsilon_{s'}(\bm p')\cdot\bm k&=\frac{\bm k^4\bm p^4\left(1-\frac{\bm k^2}{4\bm p^2}\right)^2}{\bm k^2\bm p^4\left(1-\frac{\bm k^2}{4\bm p^2}\right)}\nonumber\\
&=\bm k^2\left(1-\frac{\bm k^2}{4\bm p^2}\right)\,,
\end{align}
which is nothing more than Eq.~\eqref{eq:spin_averaged_sum_2}. Let us now perform the sum over $s$ and $s'$ in Eq.~\eqref{eq:scattering_amplitude_photon_no_sum} as follows
\begin{align}\label{eq:scattering_amplitude_photon_sum}
\sum_{s,s'}\mathcal{A}_{ss'}=\frac{\kappa^2}{\mathcal{F}(-\bm k^2)\bm k^2}\bigg\{&\left[2(p\cdot q)^2+\bm k^2p\cdot q+\frac{\bm k^2q^2}{2}\right]\sum_{s,s'}\bm\epsilon_s(\bm p)\cdot\bm\epsilon_{s'}(\bm p')\nonumber\\
&-2\left(p\cdot q+\frac{q^2}2\right)\sum_{s,s'}\bm\epsilon_s(\bm p)\cdot\bm k\ \bm\epsilon_{s'}(\bm p')\cdot\bm k\bigg\}\,.
\end{align}
Substituting in Eqs.~\eqref{eq:eps_dot_eps} and \eqref{eq:eps_dot_k} yields
\begin{align}\label{eq:scattering_amplitude_photon_sum_substitution}
\sum_{s,s'}\mathcal{A}_{ss'}=\frac{\kappa^2}{\mathcal{F}(-\bm k^2)\bm k^2}\bigg[&2(p\cdot q)^2\left(2-\frac{\bm k^2}{2\bm p^2}\right)+\left(\bm k^2p\cdot q+\frac{\bm k^2q^2}{2}\right)\left(2-\frac{\bm k^2}{2\bm p^2}\right)\nonumber\\
&-2\left(p\cdot q+\frac{q^2}2\right)\bm k^2\left(1-\frac{\bm k^2}{4\bm p^2}\right)\bigg]\,.
\end{align}
The last two terms in \eqref{eq:scattering_amplitude_photon_sum_substitution} cancel and one is left with Eq.~\eqref{eq:scattering_amplitude_photon}.

The gravitational potential~\eqref{eq:potential_photon_general} can be derived by applying Eq.~\eqref{eq:potential_from_scattering_amplitude} to Eq.~\eqref{eq:scattering_amplitude_photon}. It is worth pointing out that, since $p^2=0$ for the photon, we may write $E_p=|\bm p|$ and thus
\begin{align}
\left(p\cdot q\right)^2=E^2_pE^2_q\left(1+\frac{2E_p}{E_q}+\frac{E_p^2}{E_q^2}\right)\,,
\end{align} 
which we make use of when writing the potential in the form given in Eq.~\eqref{eq:potential_photon_general}.
\subsection{Scattering amplitude for the case of a spin-1/2 particle and a spinless particle}\label{sec:appendix_spin}
From Eq.~\eqref{eq:spinor_field_energy_momentum_tensor_final}, we have the following
\begin{align}\label{eq:spin_V}
V_{A\mu\nu}^{(s,s')}=\frac14\bar u_{s'}(p')\left[\gamma_\mu(p_\nu+p'_\nu)+(p_\mu+p'_\mu)\gamma_\nu\right]u_s(p)\,,
\end{align}
while $V_{B\mu\nu}$ is given by Eq.~\eqref{eq:V_B_spinless}. We now compute
\begin{align}\label{eq:V_A_V_B_munu_spin}
&V_{A\mu\nu}^{(s,s')}V_B^{\mu\nu}=\frac12\bar u_{s'}(p')\gamma^\mu(2p^\nu+k^\nu)\left[2q_\mu q_\nu-k_\mu q_\nu-k_\nu q_\mu+\eta_{\mu\nu}\frac{k^2}2\right]u_s(p)\nonumber\\
&=\frac12\bar u_{s'}(p')\left[4\left(p\cdot q\right)\slashed q-2\left(p\cdot k\right)\slashed q-2\left(p\cdot q\right)\slashed k+k^2\slashed p+2\slashed q\left(q\cdot k\right)-\left(q\cdot k\right)\slashed k-k^2\slashed q+\frac{k^2}2\slashed k\right]u_s(p)\nonumber\\
&=\frac12\bar u_{s'}(p')\left[4\left(p\cdot q\right)\slashed q-2\left(p\cdot q\right)\slashed k+k^2\left(\slashed p+\slashed q\right)\right]u_s(p)\,,
\end{align}
where, after the third equality, we have made use of $q\cdot k=-p\cdot k=k^2/2$. We also need to compute $V_A^{(s,s')}V_B$ which requires Eq.~\eqref{eq:V_B_spinless_trace} and
\begin{align}
V_A^{(s,s')}=\frac12\bar u_{s'}(p')\left(2\slashed p+\slashed k\right)u_s(p)\,,
\end{align}
which can be obtained by taking the trace of Eq.~\eqref{eq:spin_V}. We now have
\begin{align}\label{eq:V_A_V_B_spin}
V_A^{(s,s')}V_B=\bar u_{s'}(p')\left(2q^2\slashed p+q^2\slashed k+k^2\slashed p+\frac{k^2}2\slashed k\right)u_s(p)\,.
\end{align}
Through the use of Eqs.~\eqref{eq:V_A_V_B_munu_spin} and \eqref{eq:V_A_V_B_spin} one can now obtain
\begin{align}\label{eq:V_A_V_B_spin_final}
V_{A\mu\nu}^{(s,s')}V_B^{\mu\nu}-\frac12V_A^{(s,s')}V_B=\frac12\bar u_{s'}(p')\left[4\slashed q\left(p\cdot q\right)-2\slashed pq^2+\slashed qk^2-\slashed k\left(2p\cdot q+q^2+\frac{k^2}2\right)\right]u_s(p)\,.
\end{align}
Substituting Eq.~\eqref{eq:V_A_V_B_spin_final} into Eq.~\eqref{eq:S_matrix_V_A_B} yields Eq.~\eqref{eq:scattering_matrix_spin_half}.

At this point, we consider the CM frame. In such a frame, the scattering amplitude for the single graviton exchange between a massive spin-1/2 particle and a massive spinless particle is
\begin{align}\label{eq:scattering_amplitude_spin_half}
\mathcal{A}_{ss'}=\frac{\kappa^2}{4\mathcal{F}(-\bm k^2)\bm k^2}\bar u_{s'}(p')\left[4\slashed qp\cdot q-2\slashed pq^2+\bm k^2\slashed q-\bm\gamma\cdot\bm k\left(2p\cdot q+q^2+\frac{\bm k^2}{2}\right)\right]u_s(p)\,,
\end{align}
which can be obtained by applying Eq.~\eqref{eq:extract_scattering_amplitude} to the $S$-matrix element~\eqref{eq:scattering_matrix_spin_half}. The explicit dependence of the scattering amplitude~\eqref{eq:scattering_amplitude_spin_half} on the graviton transfer momentum $k$ is not yet clear since the $\bar u_{s'}(p')$ depends on $k$. In order to examine such a dependence, we wish to make use of Eqs.~\eqref{eq:u_0_u} and \eqref{eq:u_i_u} which are used in order to write the $u_s(p)$ in terms of the four-momentum of the spin-1/2 particle. Before making use of these two expressions, we shall first derive them below for the sake of completeness.

As already mentioned, in the present work we make use of the gamma matrices in the Weyl representation which are given by~\cite{Peskin:1995ev}
\begin{align}\label{eq:gamma_matrices}
\gamma^0=\begin{pmatrix}
0 & \mathbb{1}_{2\times2}\\
\mathbb{1}_{2\times2} & 0
\end{pmatrix}\,,\hspace{2cm}\gamma^i=\begin{pmatrix}
0 & \sigma^i\\
-\sigma^i & 0
\end{pmatrix}\,,
\end{align}
where 
\begin{align}
\sigma^1=\begin{pmatrix}
0 & 1\\
1 & 0
\end{pmatrix}\,,\hspace{1cm}\sigma^2=\begin{pmatrix}
0 & -i\\
i & 0
\end{pmatrix}\,,\hspace{1cm}\sigma^3=\begin{pmatrix}
1 & 0\\
0 & -1
\end{pmatrix}\,,
\end{align}
are the Pauli spin matrices. By making use of $\sigma_i^2=1$ and $\sigma_i\sigma_j=-\sigma_j\sigma_i$ for $i\neq j$, we have $\left(\bm\sigma\cdot\bm p\right)^2=\bm p^2$ as well as the following expressions up to second-order in three-momentum terms
\begin{align}
\sqrt{p^0\pm\bm\sigma\cdot\bm p}&\approx\sqrt{p^0}\pm\frac{\bm\sigma\cdot\bm p}{2\sqrt{p^0}}-\frac{\bm p^2}{8(p^0)^{3/2}}\,,\\
\sqrt{p'^0\pm\bm\sigma\cdot\bm p'}&\approx\sqrt{p^0}\pm\frac{\bm\sigma\cdot(\bm p+\bm k)}{2\sqrt{p^0}}-\frac{\bm p^2}{8(p^0)^{3/2}}\,.
\end{align}
We now proceed to compute
\begin{align}
&\bar u_{s'}(p')\gamma^0u_s(p)=\xi_{s'}^\dagger\left[\sqrt{p'^0+\bm p'\cdot\bm\sigma}\sqrt{p^0+\bm\sigma\cdot\bm p}+\sqrt{p'^0-\bm p'\cdot\bm\sigma}\sqrt{p^0-\bm\sigma\cdot\bm p}\right]\xi_s\nonumber\\
&\approx\xi_{s'}^\dagger\bigg[p^0+\bm\sigma\cdot\bm p+\frac{\bm\sigma\cdot\bm k}{2}-\frac{\bm p^2}{4p^0}+\frac{(\bm\sigma\cdot\bm p)^2}{4p^0}+\frac{\left(\bm\sigma\cdot\bm k\right)\left(\bm\sigma\cdot\bm p\right)}{4p^0}+p^0-\bm\sigma\cdot\bm p-\frac{\bm\sigma\cdot\bm k}{2}\nonumber\\
&\hspace{4cm}-\frac{\bm p^2}{4p^0}+\frac{(\bm\sigma\cdot\bm p)^2}{4p^0}+\frac{\left(\bm\sigma\cdot\bm k\right)\left(\bm\sigma\cdot\bm p\right)}{4p^0}\bigg]\xi_s\,,
\end{align}
which simplifies to Eq.~\eqref{eq:u_0_u}. We now turn out attention to considering
\begin{align}
&\bar u_{s'}(p')\gamma^iu_s(p)=\xi_{s'}^\dagger\left[\sqrt{p'^0+\bm p'\cdot\bm\sigma}\sigma^i\sqrt{p^0+\bm\sigma\cdot\bm p}-\sqrt{p'^0-\bm p'\cdot\bm\sigma}\sigma^i\sqrt{p^0-\bm\sigma\cdot\bm p}\right]\xi_s\nonumber\\
&\approx\xi_{s'}^\dagger\bigg[p^0\sigma^i+\frac{\sigma^i\bm\sigma\cdot\bm p}2+\frac{\bm\sigma\cdot(\bm p+\bm k)\sigma^i}2-\frac{\bm p^2\sigma^i}{4p^0}+\frac{\bm\sigma\cdot(\bm p+\bm k)\sigma^i\bm\sigma\cdot\bm p}{4p^0}-p^0\sigma^i+\frac{\sigma^i\bm\sigma\cdot\bm p}2\nonumber\\
&\hspace{4cm}+\frac{\bm\sigma\cdot(\bm p+\bm k)\sigma^i}2+\frac{\bm p^2\sigma^i}{4p^0}-\frac{\bm\sigma\cdot(\bm p+\bm k)\sigma^i\bm\sigma\cdot\bm p}{4p^0}\bigg]\xi_s\,,\nonumber\\
&=\xi_{s'}^\dagger\bigg[\sigma^i\bm\sigma\cdot\bm p+\bm\sigma\cdot\bm p\sigma^i+\bm\sigma\cdot\bm k\sigma^i\bigg]\xi_s\,.\label{eq:u_i_u_derive}
\end{align}
By making use of $\sigma^i\bm\sigma\cdot\bm p+\bm\sigma\cdot\bm p\sigma^i=2p^i$, Eq.~\eqref{eq:u_i_u_derive} reduces to Eq.~\eqref{eq:u_i_u}.

Let us now make use of Eqs.~\eqref{eq:u_0_u} and \eqref{eq:u_i_u} in the scattering amplitude~\eqref{eq:scattering_amplitude_spin_half} and obtain
\begin{align}\label{eq:scattering_amplitude_spin_half_2}
\mathcal{A}_{ss'}\approx\frac{\kappa^2\xi_{s'}^\dagger}{4\mathcal{F}(-\bm k^2)\bm k^2}\bigg\{\left(-4q^0p\cdot q+2p^0q^2-\bm k^2q^0\right)\left[2p^0+\frac{(\bm\sigma\cdot\bm k)(\bm\sigma\cdot\bm p)}{2p^0}\right]\nonumber\\
+\left(4q^ip\cdot q-2p^iq^2+\bm k^2q^i\right)\left(2p_i+\bm\sigma\cdot\bm k\sigma_i\right)\bigg\}\xi_s\,.
\end{align}
It is worth pointing out that the $\bm\gamma\cdot\bm k$ term appearing in Eq.~\eqref{eq:scattering_amplitude_spin_half} does not contribute since
\begin{align}
&\bar u_{s'}(p')\bm\gamma\cdot\bm ku_s(p)\approx\xi^\dagger_{s'}\left[2\bm p\cdot\bm k+(\bm\sigma\cdot\bm k)^2\right]\xi_s=0\,,
\end{align}
which is zero since $\left(\bm\sigma\cdot\bm k\right)^2=\bm k^2$ and $\bm p\cdot\bm k=-\bm k^2/2$. We now expand out Eq.~\eqref{eq:scattering_amplitude_spin_half_2} and use $\bm q=-\bm p$ to obtain
\begin{align}\label{eq:scattering_amplitude_spin_half_3}
\mathcal{A}_{ss'}\approx\frac{\kappa^2\xi_{s'}^\dagger}{4\mathcal{F}(-\bm k^2)\bm k^2}\bigg\{8\left(p\cdot q\right)^2-4p^2q^2-2p^0q^0\bm k^2+(\bm\sigma\cdot\bm k)(\bm\sigma\cdot\bm p)\left[(2+2-1)(q^0)^2+4p^0q^0\right]\bigg\}\xi_s\,,
\end{align}
and thus, after using $\left(\bm\sigma\cdot\bm k\right)\left(\bm\sigma\cdot\bm p\right)=i\bm\sigma\cdot(\bm k\times\bm p)-\bm k^2/2$, find
\begin{align}\label{eq:scattering_amplitude_spin_half_4}
\mathcal{A}_{ss'}\approx\frac{\kappa^2\xi_{s'}^\dagger}{4\mathcal{F}(-\bm k^2)\bm k^2}\bigg\{8\left(p\cdot q\right)^2-4p^2q^2-\left[\left(2+\frac42\right)p^0q^0+\frac32\left(q^0\right)^2\right]\bm k^2\nonumber\\
+i\bm\sigma\cdot(\bm k\times\bm p)\left[3(q^0)^2+4p^0q^0\right]\bigg\}\xi_s\,.
\end{align}
Performing the summation of Eq.~\eqref{eq:scattering_amplitude_spin_half_4} over $s$ and $s'$ yields Eq.~\eqref{eq:sum_scattering_amplitude_spin}.

As already mentioned, for the interaction between a spin-1/2 particle and a spinless particle, we consider the non-relativistic approximation and therefore consider the potential up to second-order in three-momentum terms. When deriving the potential through the use of Eq.~\eqref{eq:potential_from_scattering_amplitude}, the scattering amplitude is divided by $E_pE_q=p^0q^0$. Dividing Eq.~\eqref{eq:sum_scattering_amplitude_spin} by such a factor yields
\begin{align}\label{eq:sum_scattering_amplitude_spin_divide_p0_q0}
\frac1{p^0q^0}\left(\frac12\sum_{s,s'}\mathcal{A}_{ss'}\right)\approx\frac{\kappa^2\psi^\dagger}{\mathcal{F}(-\bm k^2)\bm k^2}\bigg[\frac{2\left(p\cdot q\right)^2}{p^0q^0}&-\frac{p^2q^2}{p^0q^0}+i\bm\sigma\cdot\left(\bm k\times\bm p\right)\left(1+\frac{3q^0}{4p^0}\right)\nonumber\\
&-\bm k^2\left(1+\frac{3q^0}{8p^0}\right)\bigg]\psi\,.
\end{align}
We now wish to write Eq.~\eqref{eq:sum_scattering_amplitude_spin_divide_p0_q0} explicitly in terms of the $A$ and $B$ particles' three-momentum $\bm p$ given that we are considering the non-relativistic approximation. To this end, let us first make use of Eq.~\eqref{eq:p_dot_q_squared} and write
\begin{align}\label{eq:divide_p_dot_q}
\frac{\left(p\cdot q\right)^2}{p^0q^0}=E_pE_q+2\bm p^2+\frac{\bm p^4}{E_pE_q}\,.
\end{align}
Using $E_p\approx m_A+\bm p^2/(2m_A)$ and similarly $E_q\approx m_B+\bm p^2/(2m_B)$, Eq.~\eqref{eq:divide_p_dot_q} gives the following up to second-order in $\bm p$
\begin{align}\label{eq:expand_p_dot_q}
\frac{\left(p\cdot q\right)^2}{p^0q^0}\approx m_Am_B+\left(\frac{m_B}{2m_A}+\frac{m_A}{2m_B}\right)\bm p^2+2\bm p^2\,.
\end{align}
Let us now turn our attention to Eq.~\eqref{eq:p_q_squared} which gives the following after dividing by $p^0q^0$
\begin{align}
\frac{p^2q^2}{p^0q^0}=E_pE_q-\left(\frac{E_p}{E_q}+\frac{E_q}{E_p}\right)\bm p^2+\frac{\bm p^4}{E_pE_q}\,,
\end{align}
which, when expanded up to second-order in three-momentum terms, gives
\begin{align}\label{eq:expand_p_q_squared}
\frac{p^2q^2}{p^0q^0}\approx m_Am_B+\left(\frac12-1\right)\left(\frac{m_A}{m_B}+\frac{m_B}{m_A}\right)\bm p^2\,.
\end{align}
Thus, by making use of Eqs.~\eqref{eq:expand_p_dot_q} and~\eqref{eq:expand_p_q_squared}, we have
\begin{align}\label{eq:expand_p_q_final}
\frac{2\left(p\cdot q\right)^2}{p^0q^0}-\frac{p^2q^2}{p^0q^0}\approx m_Am_B+\left(4+\frac{3m_A}{2m_B}+\frac{3m_B}{2m_A}\right)\bm p^2\,.
\end{align}
Let us now turn our attention to the remaining terms in Eq.~\eqref{eq:sum_scattering_amplitude_spin_divide_p0_q0}. The $i\bm\sigma\cdot(\bm k\times\bm p)$ and $\bm k^2$ factors are already second-order in three-momenta. We therefore approximate $q^0/p^0\approx m_B/m_A$ in the last two terms in Eq.~\eqref{eq:sum_scattering_amplitude_spin_divide_p0_q0} which, together with Eq.~\eqref{eq:expand_p_q_final}, yields
\begin{align}\label{eq:sum_scattering_amplitude_spin_divide_p0_q0_approx}
-\frac1{4p^0q^0}\left(\frac12\sum_{s,s'}\mathcal{A}_{ss'}\right)&\approx-\frac{\kappa^2\psi^\dagger}{4\mathcal{F}(-\bm k^2)\bm k^2}\bigg[m_Am_B+\left(4+\frac{3m_A}{2m_B}+\frac{3m_B}{2m_A}\right)\bm p^2\nonumber\\
&+i\bm\sigma\cdot\left(\bm k\times\bm p\right)\left(1+\frac{3m_B}{4m_A}\right)-\bm k^2\left(1+\frac{3m_B}{8m_A}\right)\bigg]\psi\,.
\end{align}
By taking the inverse Fourier transform of Eq.~\eqref{eq:sum_scattering_amplitude_spin_divide_p0_q0_approx} and introducing $\hbar$ and $c$ through dimensional analysis, one arrives at $\psi^\dagger V(\bm r)\psi$ where $V(\bm r)$ is the gravitational potential given in Eq.~\eqref{eq:spin_1/2_potential}.

\end{document}